\documentclass[12pt,preprint]{aastex}
\usepackage{emulateapj5}

\begin{document}

\title{Cosmic evolution of dust in galaxies: Methods and
preliminary results}

\author{Kenji Bekki} 
\affil{
ICRAR,
M468,
The University of Western Australia
35 Stirling Highway, Crawley
Western Australia, 6009, Australia
}

\begin{abstract}

We investigate the redshift ($z$) evolution of dust mass and abundance,
their  dependences on initial conditions of galaxy formation,
and
physical correlations between dust, gas, and stellar contents at different $z$
based on our original chemodynamical simulations of galaxy formation
with dust growth and destruction.
In this preliminary investigation, we first determine the reasonable ranges
of the most important two parameters for dust evolution, i.e.,
the timescales of dust growth and destruction, by comparing
the observed and simulated
dust mass and abundances and  molecular hydrogen (${\rm H_2}$) content
of the Galaxy. We then investigate the $z$-evolution
of dust-to-gas-ratios ($D$)
and, ${\rm H}_2$ gas fraction ($f_{\rm H_2}$),
and gas-phase chemical abundances (e.g., $A_{\rm O}=12+\log({\rm O/H})$)
in the simulated disk and dwarf galaxies.
The principal results are as follows.
Both $D$ and $f_{\rm H_2}$ can rapidly increase during the early dissipative formation
of galactic disks ($z \sim 2-3$) and the $z$-evolution of these depends on
initial mass densities, spin parameters, and masses of galaxies.
The observed $A_{\rm O}-D$ relation can be qualitatively reproduced, but the simulated
dispersion of $D$ at a given $A_{\rm O}$ is smaller. The simulated galaxies
with larger total dust masses show larger ${\rm H_2}$ and stellar masses and
higher $f_{\rm H_2}$. Disk galaxies show negative radial gradients
of $D$ and the gradients are steeper for more massive galaxies.
The observed evolution of dust masses and dust-to-stellar-mass-ratios
between $z=0$ and $0.4$ can not be reproduced so well by the simulated disks.
Very extended dusty gaseous
halos can be formed during hierarchical build-up of disk galaxies.
Dust-to-metal ratios  (i.e., dust-depletion levels)
are different within a single galaxy and between different galaxies
at different $z$.

\end{abstract}

\section{Introduction}

Interstellar dust plays a number of key roles in galaxy formation and evolution. 
For example,
molecular hydrogen (${\rm H_2}$) can form on the surface of dust grains
(e.g., Gould \& Salpeter 1963; Hollenbach \& Salpeter 1971) so that 
the time evolution of ${\rm H_2}$ content and the
star formation processes in giant molecular clouds can be influenced by
dust abundances in galaxies.
Dust grains are fundamentally important for the fragmentation processes
of metal-poor gas clouds and thus for the formation of low-mass stars in
the early universe (e.g., Larson 2005; Schneider \& Omukai 2010).
Spectral energy distributions (SEDs) of galaxies are strongly influenced
by dust abundances,  compositions, and size-distributions within galaxies
(e.g., Popescu et al. 2000; Takeuchi et al. 2005).
Therefore,
the time evolution of photometric and spectroscopic properties,
star formation rates, and cold gas contents in galaxies can be 
strongly
coupled with the cosmic evolution of interstellar dust.
Theoretical studies have not yet investigated
how interstellar dust can possibly influence galaxy formation and evolution
processes in detail.

A growing number of observational studies by space infrared telescopes 
(e.g., Spitzer, Herschel, and AKARI) have revealed  dust properties for
a large number of galaxies with different masses
and types in different environments and their correlations with
galaxy properties such as total stellar masses (e.g. Draine et al. 2007;
Dunne et al. 2011, D11; Kaneda et al. 2011; Cortese et al. 2012;
Smith et al. 2012; Davies et al. 2014).
Furthermore,  these observations
have also revealed the spatially resolved dust properties in nearby galaxies
and accordingly provided new constraints on dust evolution models in galaxies 
(e.g., Meixner et al. 2010; Pappalardo et al. 2012; P12).
In spite of these significant progresses in observational studies,
theoretical studies have not yet explained many of these latest new observational
results.

Physical properties of dust have been recently observed to be different between
galaxies at different $z$.
For example,
one of the most
distant quasars, SDSS J1148+5251 at $z=6.4$,  has been observed to have a large amount
of ${\rm H_2}$ gas and dust (e.g., Bertoldi et al. 2003; Beelen et al. 2006).
Recent ALMA observations of
a very high redshift star-forming galaxy at $z=6.6$ ('Himiko'), on the other hand,
revealed that this galaxy has a very low dust content consistent with primordial 
interstellar gas (Ouchi et al. 2013).
Kashino et al. (2013) have investigated the level of dust extinction in distant
galaxies at $1.4 < z < 1.7$ by using the results of a near-IR spectroscopic
survey of the COSMOS field and found that the extinction level is elevated
at high mass galaxies compared to low-$z$ counterparts.
These observed differences of dust properties in galaxies at different $z$  
and the $z$-evolution of galactic dust properties  are yet to be explained
by theoretical models of galaxy formation.

The time evolution of dust and physical correlations between dust properties and 
other galaxy properties (e.g., gas-phase abundances) have long been investigated
mainly by one-zone or multi-zone  chemical evolution models
(e.g., Dwek \& Scalo 1979, 1980;
Dwek 1998, D98; Lisenfeld \& Ferrara 1998; Hirashita 1999; Edmunds 2001;
Inoue 2003;
Calura et al. 2008; Asano et al. 2013; Zhukovska \& Henning 2013). 
Owing to the lack of spatial resolution in these models,
2D distributions of dust properties in galaxies were not investigated in the models,
though they already discussed a few dust scaling relation (e.g., $A_{\rm O}-D$ 
correlation).  
The $z$-evolution of dust properties in galaxies was not investigated in these
previous one-zone models that are not based on a cosmological model.
Chemodynamical simulations with dust evolution could be ideal theoretical tools
for discussing the spatially resolved dust properties of galaxies and 
redshift evolution of dust revealed in recent observations described above.

Bekki (2013a, B13a) first incorporated both dust growth and destruction and
${\rm H_2}$ formation on dust grains
in the chemodynamical simulations with star formation from ${\rm H_2}$ gas
and thereby investigated the time evolution of dust and ${\rm H_2}$
properties of galaxies in a  self-consistent manner.
However, the initial conditions of galaxy formation in B13a
are highly idealized for its preliminary investigation of
dust evolution and are not those adopted in numerical simulations of galaxy formation
based on a $\Lambda$CDM cosmology.
Therefore, neither the $z$-evolution of dust properties 
nor the roles of hierarchical growth of galaxies in dust evolution
were investigated in B13a. Given a growing number of observational
studies on dust properties of galaxies at different $z$,
it is timely for theoretical studies of galaxy formation to investigate
the evolution of dust in the context of hierarchical galaxy formation.

The purpose of this paper is to investigate the $z$-evolution
of dust  and ${\rm H_2}$ properties of galaxies
and correlations between dust and galaxy properties 
(e.g., total stellar masses) at $z=0$  by using
numerical simulations of galaxy formation based on $\Lambda$CDM cosmology.
We particularly investigate (i) whether the observed $A_{\rm O}-D$ relation
can be reproduced by the present simulations,
(ii) how the $z$-evolution of $D$ and $f_{\rm H_2}$ (dust-to-metal-ratio) depends on
initial masses and spin parameters of galaxies,
and (iii) how $D$ and $f_{\rm H_2}$ 
are spatially different
within a galaxy,
(iv) whether the observed dust scaling relation can be reproduced,
and (v) how dust evolution is different between luminous disk galaxies
and dwarfs. 
The present study also predicts physical correlations between dust,  ${\rm H_2}$, 
and stellar 
properties of galaxies in order to compare the predictions with the corresponding
observations 
(e.g., Norman \& Spaans 1997;
Leroy et al. 2011; Saintonge et al. 2011; Cortese et al. 2012;
Bourne et al. 2013).
In this investigation, we use a slightly revised version of our previous
simulation code used in B13a.

The plan of the paper is as follows: In the next section,
we describe initial conditions of galaxy formation and
chemodynamical model with the formation and evolution
of dust and ${\rm H}_2$.
In \S 3, we
present the numerical results
on the $z$-evolution of dust and ${\rm H_2}$ properties 
in simulated  galaxies.
In this section, we also discuss correlations between dust properties
(e.g., $D$) and global galaxy parameters (e.g., total stellar masses)
and their $z$-evolution.
In \S 4, we discuss the latest observational results on dust and ${\rm H}_2$
properties of galaxies derived mainly from the $Herschel$ and $Spitzer$ telescopes
and also point out necessary improvement in the predictive power of
numerical simulations with dust growth and destruction.
We summarize our  conclusions in \S 5.
Physical meanings for symbols (e.g., $D$) often used in the present study
are given in Table 1 for convenience.

\section{The model}

We mainly investigate the $z-$evolution of total dust and ${\rm H_2}$ masses
($M_{\rm dust}$ and $M_{\rm H_2}$, respectively),
$D$, mass fraction of PAH dust ($q_{\rm PAH}$) from $z \sim 16$ to $z=0$ by using
chemodynamical simulations of galaxy formation based on a $\Lambda$CDM cosmology
with the latest WMAP cosmological parameters. Since the details
of the simulation code adopted
in the present study and the results of various tests of the code have been
already given in B13a, we here briefly describe the simulation code.
This study presents the first application of the new simulation 
code to galaxy formation in a cosmological context.
Accordingly it is still a preliminary one in which (i) a reasonable range of
dust parameters needs to be determined in the context of CDM-based galaxy formation 
and (ii) a large-scale structure and formation processes of individual
galaxies are not self-consistently included (i.e., not 'zoom-in' simulations).

Although a growing number of observational papers on dust properties
of local and distant galaxies have been published,  many observational studies
are currently underway.
Therefore,
only a brief comparison between observations and simulations can be done in the present
study.
Ongoing data analysis from Herschel and ALMA will produce a large amount of
key observational data that can be compared with numerical simulations
of galaxy formation with dust and ${\rm H_2}$ models.
It is thus our future work to compare extensively 
between more sophisticated simulations and
these observations.

We use a slightly revised version of our previous simulation code (B13a) in which
hydrodynamics of interstellar gas is modeled by the SPH method.
The previous simulations in B13a
include the formation of dust in the stellar winds of supernovae (SNe)
and asymptotic giant branch (AGB) stars, the growth and destruction processes
of dust in the interstellar
medium (ISM),  the formation of polycyclic aromatic hydrocarbon (PAH) dust
in carbon-rich AGB stars,
the ${\rm H_2}$ formation on dust grains,
and the ${\rm H_2}$ photo-dissociation due to far ultra-violet (FUV)
light  in a self-consistent manner. The revised version newly includes
(i) dust growth timescales dependent on dust properties (Yozin \& Bekki 2014) and 
(ii) time-varying IMF models (Bekki \& Meurer 2013; 
Bekki 2013b,c,  B13b, B13c). These new elements of the code
are described later.
The code can be currently run on Graphics Processing Unit (GPU)
clusters  in which gravitational calculations
are done on GPUs whereas other calculations such as star formation and chemical evolution
are done on CPUs.

The present improved code, however, does not allow us to investigate the following
physical properties of dust. First, the time evolution of dust size distribution
and detailed dust compositions can not be investigated. This is because initial
distributions of dust sizes and dust shattering and coagulation processes,
which are key for the evolution of dust grain sizes, are not included.
Second, the dynamical evolution of dust can not be separately investigated
from that of gas-phase metals (i.e., metals that are not locked up onto dust grains),
because dust-gas hydrodynamical interaction (e.g. Theis \& Orlova 2004)
and radiation pressure on dust grains (e.g., Ferrara et al. 1991)
are not included.
Third, the sub-pc scale physical processes of dust destruction  by  SNe
and grain-grain collisions 
(e.g., Shull 1978;
McKee et al. 1989; Jones et al. 1994; Silvia et al. 2010)  can not be investigated by
the present galaxy-scale
simulations because of the expected spatial resolution (at most 10pc).
These three will be included in our future more sophisticated simulations.

As shown later in this paper, the simulated disk galaxies can have extended
dusty gaseous halos. Therefore, the total mass of dust in a  simulated galaxy
depends on the radius within which the masses of gas particles
are summed up. A significant fraction of dust in a galaxy can be
located in the outer halo. Therefore, we separately estimate
the total dust dust mass for disk and halo components of a galaxy.
In the present study,  $M_{\rm dust}$,
$M_{\rm dust,d}$, and $M_{\rm dust, h}$
are referred to  as  the total masses of 
a galaxy,  that of the disk ($R<R_{0.5}$, where $R_{0.5}$ is the
half-mass radius of stars),  
and that of the halo, respectively.

\subsection{Gravitational dynamics and hydrodynamics}

One of key ingredients of the code is that
the gravitational softening length ($\epsilon$) is chosen for each
component in a galaxy (i.e.,
multiple gravitational softening lengths).
Thus the gravitational softening length ($\epsilon$)
is different between dark matter ($\epsilon_{\rm dm}$)
and gas ($\epsilon_{\rm g}$) and $\epsilon_{\rm dm}$ is determined by the initial
mean separation of dark matter particles.
Initial $\epsilon_{\rm g}$ is set to be significantly smaller than 
$\epsilon_{\rm dm}$ owing to rather high-density (thus high-number-density) gaseous regions
formed during dissipative collapse and merging of galaxy formation, as is described later.
The softening length for new stars formed from gas is set to be the same 
as $\epsilon_{\rm g}$.
Furthermore,  when two different components interact gravitationally,
the mean softening length for the two components
is applied for the gravitational calculation.
For example, $\epsilon = ({\epsilon}_{\rm dm}+{\epsilon}_{\rm g})/2$
is used for gravitational interaction between

We consider that the ISM in galaxies can be modeled as an ideal gas with
the ratio of specific heats ($\gamma$) being 5/3.
The basic methods to implement SPH in the present study are essentially
the same as those proposed by Hernquist \& Katz (1989).
We adopt the predictor-corrector algorithm (that is accurate to second order
in time and space) in order to integrate the equations
describing  the time  evolution of a system.
Each particle is allocated an individual time step width ($\Delta t$) that is determined
by physical properties of the particle.
 The maximum time step width ($\Delta t_{\rm max}$)
is $0.01$ in simulation units, which means that  $\Delta t_{\rm max}=1.41 \times 10^6$ yr
in the present study. Although a gas particle is allowed to have a minimum time step
width of $1.41 \times 10^4$ in the adopted individual time step scheme,
no particle actually has such a short time step width owing to conversion
from gas to star in high-density gas regions.
The radiative cooling processes
are properly included  by using the cooling curve by
Rosen \& Bregman (1995) for  $100 \le T_{\rm g} < 10^4$K
(where $T_{\rm g}$ is gas temperature)
and the MAPPING III code
for $T_{\rm g} \ge 10^4$K
(Sutherland \& Dopita 1993).

In the present study,  thermal feedback effects of SN explosions can heat up
ISM to form hot gas whereas radiative cooling described above can contribute
to the formation of rather cold gas ($T_{\rm g} \sim 100$K). 
Therefore, the influences of the multi-phase nature of ISM on dust evolution
can be investigated, though the details of ISM evolution driven by
gas cooling and heating, SNe, and  ${\rm H_2}$ formation on dust grains
are not discussed in the present paper.
Also, dust evolution can be influenced by stellar winds of massive stars
and proto-stellar objects. These potentially important processes for dust
evolution need to be investigated in our future papers, because the present
simulations do not have enough resolution to investigate this subpc-scale
dynamics. The cosmic microwave background (CMB) radiation field can also
influence the evolution of cold gas and thus that of
interstellar dust  in galaxies at higher redshifts.
This CMB effects on ISM with dust are not considered in the present
study, and therefore it 
is also a key issue that should be investigated in our forthcoming papers.

\subsection{Chemical enrichment}

Since the present model for chemical enrichment processes of galaxies
is exactly the same as that used in B13a, we briefly describe the
model here.
Chemical enrichment through star formation and metal ejection from
SNIa, II, and AGB stars is self-consistently included in the chemodynamical
simulations.
We investigate the time evolution of the 11 chemical elements of H, He, C, N, O, Fe,
Mg, Ca, Si, S, and Ba in order to predict both chemical abundances and dust properties
in the present study. The mean metallicity $Z$ for each $k$th stellar particle is
represented by $Z_k$. The total mass of each $j$th ($j=1-11$) chemical component
ejected from each $k$th stellar particles at time $t$ is given as
\begin{equation}
\Delta z_{k,j}^{\rm ej}(t)=m_{\rm s, \it k} Y_{k,j}(t-t_k),
\end{equation}
where $m_{\rm s, \it k}$ is the mass of the $k$th stellar particle, $Y_{k,j}(t-t_k)$
is the mass of each $j$th chemical component ejected from stars per unit mass at
time $t$, and $t_k$ represents the time when the $k$th stellar particle is
born from a gas particle. $\Delta z_{k,j}^{\rm ej}(t)$ is given equally  to neighbor SPH gas
particles (with the total number of $N_{\rm nei, \it  k}$)
located around the $k$th stellar particle.  Therefore,
the mass increase of each $j$th chemical component for $i$th gas particle at time $t$
($\Delta z_{i,j}^{\rm ej}(t)$) is
given as
\begin{equation}
\Delta z_{i,j}^{\rm ej}(t) = \sum_{k=1}^{N_{\rm nei, \it i}}
m_{\rm s, \it k} Y_{k,j}(t-t_k)/N_{\rm nei, \it k},
\end{equation}
where $N_{\rm nei, \it i}$ is the total number of neighbor stellar particles whose metals
can be incorporated into the $i$th gas particle.
We consider the time delay between the epoch of star formation
and those  of supernova explosions and commencement of AGB phases (i.e.,
non-instantaneous recycling of chemical elements).

We adopt the `prompt SN Ia' model in which
the delay time distribution (DTD)
of SNe Ia is consistent with  recent observational results by  extensive SN Ia surveys
(e.g.,  Mannucci et al. 2006).
In this prompt SN Ia mode,
there is a time delay ($t_{\rm Ia}$) between the star formation
and the metal ejection for SNe Ia. We here adopt the following DTD
($g(t_{\rm Ia}$)) for 0.1 Gyr $\le t_{\rm Ia} \le$ 10 Gyr,
which is consistent with recent observational studies
on the SN Ia rate in extra-galaxies (e.g., Maoz et al.,
2011):
\begin{equation}
g_{\rm Ia} (t_{\rm Ia})  = C_{\rm g}t_{\rm Ia}^{-1},
\end{equation}
where $C_{\rm g}$ is a normalization constant that is determined by
the number of SN Ia per unit mass  (which is controlled by the IMF
and the binary fraction for intermediate-mass stars
for  the adopted power-law slope of $-1$).
This adoption is pointed out to be necessary to explain the observed chemical properties
of the LMC (Bekki \& Tsujimoto 2012)
The fraction of the stars that eventually
produce SNe Ia for 3--8$M_{\odot}$ has not been observationally determined.
In  the present study,  $f_{\rm b}=0.05$ is assumed.
We adopt the nucleosynthesis yields of SNe II and Ia from Tsujimoto et al. (1995; T95)
and AGB stars from van den Hoek \& Groenewegen (1997; VG97)
in order to estimate $Y_{k,j}(t-t_k)$ in the present study.

\subsection{Dust model}

The present models for dust yields, growth, and destruction
are the same as those adopted in  B13a,
which reproduced reasonably well the observed
dust properties of galaxies  in a self-consistent manner.
In B13a and the present work, the evolution of dust size distribution
is not considered at all,  though it can influence the thermal evolution of ISM
and ${\rm H_2}$ formation efficiency on dust grains.
The total mass of $j$th component ($j$=C, O, Mg, Si, S, Ca, and Fe)
of dust from $k$th type of stars ($k$ = I, II, and AGB for SNe Ia, SNe II, and
AGB stars, respectively) is described as follows;
\begin{equation}
m_{\rm dust, \it j}^k= \delta_{\rm c, \it j}^k F_{\rm ej}(m_{\rm ej, \it j}^k),
\end{equation}
where $\delta_{\rm c, \it, j}^k$ is the condensation efficiency (i.e., the mass
fraction of metals that are locked up in dust grains) for each $j$th chemical component
from $k$th stellar type,
$F_{\rm ej}$ is the function that determines the total mass of metals that can be used
for dust formation,
and $m_{\rm ej, j}^k$ is the mass of $j$th component
ejected from $k$th stellar type.
The total mass of stellar ejecta is estimated by using stellar yield 
tables by T95 and VG97. Dust yields are exactly the same as those in B13a.
The details of  $\delta_{\rm c, \it j}^k$ and $  F_{\rm ej}(m_{\rm ej, \it j}^k)$
are given in B13a.
The inclusion of SN dust in the present model is quite reasonable,
firstly because recent observations of SN 1987a by Herschel have revealed
a large amount of dust ([0.4$-$0.7]${\it M}_{\odot}$) in the SN
(Matsuura et al. 2011), and secondly because
the observed amount appears to be  consistent with the theoretical prediction
by Nozawa et al. (2003).

Dust grains can grow by accretion of metals of ISM onto preexisting cores and this
accretion process is included in previous models (D98). Following D98, we consider
that the key parameter in dust accretion is the dust accretion timescale ($\tau_{\rm a}$).
In the present study, this parameter can vary between different gas particles
and is thus represented by $\tau_{\rm a, \it i}$ for $i$th gas particle.
The mass of $j$th component
($j$=C, O, Mg, Si, S, Ca, and Fe) of dust for $i$th gas particle
at time $t$ ($d_{i,j}(t)$) can increase owing  to dust accretion processes.
The mass increase
is described as
\begin{equation}
\Delta d_{i,j}^{\rm acc}(t)=\Delta t_i (1-f_{\rm dust,\it i, j})
d_{i,j}(t) /\tau_{\rm a, \it i},
\end{equation}
where $\Delta t_i$ is the individual time step width for the $i$th gas particle
and $f_{\rm dust, \it i, j}$ is the fraction of the $j$th chemical element that
is locked up in the dust. Owing to this dust growth, the mass of $j$th chemical
component that is {\it not} locked up in the dust ($z_{i,j}(t)$)
can decrease, which is simply given as
\begin{equation}
\Delta z_{i,j}^{\rm acc}(t)=- \Delta t_i (1-f_{\rm dust,\it i, j})
d_{i,j}(t) /\tau_{\rm a, \it i}
\end{equation}
As is clear in these equations, the total  mass of $j$th component in $i$th gas
particle ($m_{i,j}(t)$) is $z_{i,j}(t)+d_{i,j}(t)$.

Dust grains can be destroyed though supernova blast waves
in the ISM of galaxies (e.g., McKee 1989)
and the destruction process is parameterized by the destruction time scale
($\tau_{\rm d}$) in previous one-zone models (e.g., Lisenfeld \& Ferrara 1998;
Hirashita 1999).  Following the previous models,
the decrease  of the mass of $j$th component
of dust for $i$th gas particle
at time $t$ due to dust destruction process
is as follows
\begin{equation}
\Delta d_{i,j}^{\rm dest}(t)= - \Delta t_i
d_{i,j}(t) /\tau_{\rm d, \it i},
\end{equation}
where $\tau_{\rm d, \it i}$ is the dust destruction timescale for $i$th particle.
The dust destroyed by supernova explosions can be returned back to the ISM,
and therefore the  mass
of $j$th chemical
component that is not locked up in the dust
increases  as follows:
\begin{equation}
\Delta z_{i,j}^{\rm dest}(t)= \Delta t_i
d_{i,j}(t) /\tau_{\rm d, \it i}
\end{equation}

Thus the equation for the time evolution of $j$th component of metals
for $i$th gas particle  are given as
\begin{equation}
z_{i,j}(t+\Delta t_i)=z_{i,j}(t)+\Delta z_{i,j}^{\rm ej}(t)+\Delta z_{i,j}^{\rm acc}(t)
+\Delta z_{i,j}^{\rm dest}(t)
\end{equation}
Likewise, the equation for dust evolution is given as
\begin{equation}
d_{i,j}(t+\Delta t_i)=d_{i,j}(t)+\Delta d_{i,j}^{\rm acc}(t)
+\Delta d_{i,j}^{\rm dest}(t)
\end{equation}
Dust is locked up in stars as metals are done so, when gas particles are converted into
new stars. This means that star formation process itself has an effect
of destroying dust in the present study.

A growing number of observational studies on PAH properties have been accumulated
for galaxies within and beyond the Local Group (e.g, Draine et al. 2007; 
Meixner et al. 2010).
It is accordingly timely for the present study to discuss the origin of the observed
PAH properties in galaxies by using the new chemodynamical model.
The most  promising formation site of interstellar PAH dust is C-rich AGB stars,
though direct observation supporting PAH formation in stellar winds of AGB stars
is very weak (e.g., Tielens 2008 for a recent review).
We consider that some fraction of carbon dust produced by C-rich AGB stars ($C/O>1$)
can finally become PAH dust and thereby investigate the PAH properties in the present
study. The mass fraction of PAH dust to total carbon dust in the ejecta of C-rich AGB stars
is a parameter denoted as $R_{\rm PAH}$.
Using low-resolution simulations with different $R_{\rm PAH}$,
we find that $R_{\rm PAH}=0.05$ can better explain observations.
We therefore show the results of the models with
$R_{\rm PAH}=0.05$ in the present study.
As shown in B13a,  the time evolution of PAH is not different between
low- and high-resolution simulations for $R_{\rm PAH}=0.05$.

We adopt an assumption that the dust accretion and dust destruction timescales
are the same between PAH and other dust. Therefore, the parameter values
of $\tau_{\rm a}$ and $\tau_{\rm d}$ are exactly the same between PAH and other dust.
We adopt this assumption, mainly because it is observationally and theoretically
unclear how different the dust growth and destruction processes are between PAH
and other dust. 
It is likely, however, that PAH can be destroyed (probably more efficiently)
by gas shocks and SNe than other dust owing to their much lower masses.
This difference in dust destruction processes between different dust gains
should be included in our future more sophisticated models.
We accordingly admit that the PAH model is over-simplified to some extent,
but we consider that the PAH evolution does not influence the time evolution
of $M_{\rm dust}$ and $D$ that 
are the main focus of this paper (because $M_{\rm PAH}$ is much smaller than
$M_{\rm dust}$). In the present paper, we do not want to introduce additional few
parameters to model the PAH evolution and  more sophisticated PAH models
should be discussed in our future  papers.
The evolution of PAH dust with time and metallicity in galaxies was already
investigated in detail by Galliano et al. (2008).
Also, Micelotta et al. (2010a, b) 
investigated the effects of hot gas  and gaseous shock waves
on the evolution of PAH and estimated the lifetime and destruction process
of PAH dust.

We investigate the time evolution of the mass-ratio of PAH dust to total dust for each
$i$-th gas particle at each time step, and the mass-ratio ($q_{\rm PAH, \it i}$) 
is defined as follows:
\begin{equation}
q_{\rm PAH, \it i }=\frac{ m_{\rm PAH, \it i} } { m_{\rm dust, \it i} },
\end{equation}
where $m_{\rm dust, \it i}$ is the sum of all dust elements and $m_{\rm PAH, \it i}$
is the total mass of PAH dust for the gas particle. 
The present dust model is not so realistic as the one proposed recently by
Jones et al. (2013) which considered the evolution of hydrocarbon materials influenced
by ultraviolet photon.
We do not consider such a detailed dust model in the present study,
because the simulation code does not allow us to investigate the evolution of dust
compositions due to photon absorption-induced processing.
We will consider the new model by Jones et al. (2013) in our
future works.

\subsection{Different dust depletion levels in different elements}

Gas-phase chemical abundances relative to solar in the Galactic ISM are
observed to be quite different between different elements (e.g., Savage et al. 1992).
For example, Fe, Ca,
and Ti are severely dust-depleted (i.e., almost all Fe, Ca, and Ti are locked
up in dust) whereas N and S do not show such dust depletion. Therefore,
chemical abundance ratios, such as [S/Fe] and [S/Ca],
can be used to measure the dust depletion
level of an individual gaseous region in a galaxy.
Different dust depletion levels between different elements reflect the fact that
the required physical conditions for metals to be locked up in dust are different
between the different elements. Although the present study does not develop
a detailed model for this different dust depletion level,
it adopts a simple model in  which
only the element S is free from dust depletion.
Accordingly, the radial gradient of [S/H] in a galaxy can be regarded
as a metallicity gradient  that is not
influenced by dust depletion (thus it can correspond to the $A_{\rm O}$ 
gradient of ionized gaseous regions).
Furthermore, we can measure the dust depletion level of a galaxy 
by investigating the mean [S/Fe]
in the present study, because the element Fe is dust-depleted.

We investigate the time evolution of the dust-to-metal-ratios ($f_{\rm dust, \it i}$)
for each
$i$-th gas particle at each time step, and it 
is defined as follows:
\begin{equation}
f_{\rm dust, \it i }=\frac{ m_{\rm dust, \it i} } { m_{\rm Z, \it i} },
\end{equation}
where $m_{\rm Z, \it i}$ is the sum of all metal elements (i.e.,
metals in gas and those locked up in dust).
The mean dust-to-metal-ratio of a galaxy is the mean of these $f_{\rm dust, \it i}$
for all gas particle of the galaxy. It is observationally difficult to estimate
these $f_{\rm dust}$. Therefore, we show both [S/Fe] and $f_{\rm dust}$ in
the present study.
The initial dust-to-metal ratio is fixed at 0.1 (at $z_{\rm i}$)
for all gaseous particles in all models and the present results do not depend strongly
on this initial ratio. The initial dust and metal are assumed to originate
from Pop III SNe (e.g., Nozawa et al. 2003).

\subsection{Constant vs variable dust accretion models}

Dust growth timescale in ISM of galaxies
is a fundamentally important not only in investigating
the time evolution
of mean $D$ in galaxies (e.g., D98)  but also in discussing the physical properties
of ${\rm H_2}$ gas (B13a).
We investigate the following two models for estimating the dust growth timescale.
One is the constant dust accretion model (`CDA') in which $\tau_{\rm a}$ is constant
for all particles at all time steps and the other is the variable dust accretion
model ('VDA') in which $\tau_a$ is different between different particles
with different gaseous properties and changes with time according to the changes
of gaseous properties.
Previous one-zone models (e.g., D98) and our previous numerical simulations
(B13a) adopted CDA and thereby discussed the observed key observational results of
dust properties in galaxies. Clearly, VDA is more realist than CDA
in investigating dust and ${\rm H_2}$ properties of galaxies.

However, we need to introduce a few additional parameters in VDA in order to
describe the possible dependences of $\tau_{\rm a}$  of gas particles
on the gas densities,
temperature, and chemical abundances.
Given that the additional parameters for $\tau_{\rm a}$ are not well constrained,
VDA could be practically less useful in pinpointing the key physics for the time
evolution of dust properties. Furthermore,  our previous simulations with CDA
clearly show the importance of dust accretion and destruction in the evolution
of $D$ and $f_{\rm H_2}$. We therefore  consider that it is the best for the present
study to investigate the $z$-evolution of dust and ${\rm H_2}$ properties
of galaxies by using both CDA and VDA.

For CDA, $\tau_{\rm a, \it i}$ of each gas particle is fixed at a certain value
for all time steps and  described as follows:
\begin{equation}
\tau_{\rm a, \it i} = \tau_{\rm a},
\end{equation}
where $\tau_{\rm a}$ ranges from 0.01 Gyr to 0.5 Gyr in the present study. 
For CDA, $\tau_{\rm d, \it i}$ of each gas particle is also fixed at 
a certain value and assumed to be proportional to  $\tau_{\rm a, \it i}$ as follows:
\begin{equation}
\tau_{\rm d, \it i} = \tau_{\rm d}= \beta_{\rm d} \tau_{\rm a},
\end{equation}
where $\beta_{\rm d}$ is a free parameter and $\beta_{\rm d} \approx 2$
for $\tau_{\rm a}=0.25$ Gyr  is the
best value for reproducing dust and ${\rm H_2}$ properties of galaxies in B13a.
We can find a reasonable range of $\beta_{\rm d}$ by comparing the observed
dust and ${\rm H_2}$ properties with the simulated ones for different $\beta_{\rm d}$.

We adopt the following dependence of $\tau_{\rm a, \it i}$ on the mass density
and temperature of a gas particle:
\begin{equation}
\tau_{\rm a, \it i} = \tau_{\rm a,0}  (\frac{ \rho_{\rm g,0} }{ \rho_{\rm g, \it i} })
(\frac{ T_{\rm g,0} } { T_{\rm g, \it i} })^{0.5},
\end{equation}
where $\rho_{\rm g, \it i}$ and $T_{\rm g, \it i}$ are the gas density and temperature
of a $i$-th  gas particle, respectively,
$\rho_{\rm a,0}$ (typical ISM density at the solar neighborhood)
and $T_{\rm g,0}$ (temperature of cold gas) are set to be 1 atom cm$^{-3}$ and 20K,
respectively, and $\tau_{\rm a,0}$ is a reference dust accretion timescale
at $\rho_{\rm g,0}$ and $T_{\rm g, 0}$.
The adopted $10^8$ yr for the dust timescale of typical ISM 
is consistent with those used for typical ISM
in previous studies (e.g., $2.5 \times 10^8$ yr
in D98). The present simulations can not resolve the atomic-scale
physics of dust growth (much smaller than the possible resolution of
the present simulations), and the above $\tau_{\rm a, \it i}$ 
for a gas particle should be regarded as 
the dust growth timescale averaged over the smoothing length of the particle
($10-100$ pc). 
Furthermore,
the functional form of $\tau_{\rm a}$ in the above equation can be derived
from the first principle  (See equations 30, 31, and 34 in D98
and 19 and 20 in Hirashita 2012).

This $\tau_{\rm a,0}$ can control  the dust accretion
timescale of the $i$-th gas particle and is set to be $10^8$ yr  in the present study.
The adopted function form of $\tau_{\rm a, \it i}$ is consistent with
the analytical formula estimated by D98 (see equations 30, 31 and 34 in D98). 
Since the size and mass distributions of dust grains can determine
$\tau_{\rm a, 0}$, 
adopting a fixed value of  $\tau_{\rm a,0}$ would be less realistic. 
We cannot help but to adopt a fixed $\tau_{\rm a, 0}$ value, simply
because the present simulations do not allow us to investigate the size and
mass distributions of dust grains in galaxies.
We have investigated how the present results depend on $\tau_{\rm a,0}$ 
by running some models with longer $\tau_{\rm a,0}$.
It is confirmed that if $\tau_{\rm a,0}$ is as large as $10^9$ yr, then
the observed dust properties can not be well reproduced by the present models.
We therefore show the results of the models with $\tau_{\rm a,0} \sim 10^8$
yr.

Previous numerical simulations revealed physical conditions for dust destruction
by SNF and the dust destruction efficiencies dependent on physical properties
of ISM (e.g., Jones et al. 1994; Nozawa et al. 2003).
Although $\tau_{\rm d, \it i}$ could be described as a function of
gas properties by using the results of these simulations,
we adopt the same model in CDA for the dust destruction timescale in VDA.
This is mainly because an analytical formula for the dust destruction efficiencies
and timescales dependent on ISM properties has not been clearly described yet.
The dust destruction timescale of a gas particle depends both on $\beta_{\rm d}$
and $\tau_{\rm a, \it i}$, which means that the timescale can be determined locally
for each gas particle.  The $\beta_{\rm d}$ value is fixed at 2 for most  models
with VDA 
(unless specified)
in the present study.
In our future work, we will try to develop a better and
more sophisticated model for dust destruction
by including the results of the above previous numerical
simulations properly in our chemodynamical simulations.

The SPH smoothing length ($h$) of a particle,
which can define the resolution of a local gaseous region, is changed according to the gas
density of the particle (i.e., time-dependent, dynamic resolution).
Therefore, the $h$ value 
can be rather small  in some high-density region of a model (in particular,
where star formation is suppressed).
The density of a SPH gas particle can be therefore as high as $10^4$ cm$^{-3}$,
which is already discussed in B13a.
Furthermore, the ${\rm H_2}$
formation site is almost always in high-density (and dust-rich) regions,
which means that dust growth is the most rapid in the forming ${\rm H_2}$
regions according to the adopted
density-(temperature-)dependent dust accretion timescale.

Also, it should be stressed  that the destruction of ${\rm H_2}$
gas by supernova (or high-intensity
UV radiation field) is included already in this paper. For example, the ${\rm H_2}$
abundance is
lower in the model with strong SN feedback,  because SN feedback effects
can disperse ISM and lower the gas density so that ${\rm H_2}$ formation can be suppressed.
In such a low-density region, dust-growth is suppressed in the present simulation
(owing to the adopted density-dependent dust accretion timescale).

\subsection{${\rm H_2}$ formation and dissociation}

The model for ${\rm H_2}$ formation and dissociation in the present study
is exactly the same as those used in B13a: ${\rm H_2}$ formation
on dust grains and ${\rm H}_2$ dissociation by FUV radiation
are both self-consistently included in chemodynamical simulations.
The temperature ($T_{\rm g}$),
hydrogen density ($\rho_{\rm H}$),  dust-to-gas ratio ($D$)
of a gas particle and the strength of the
FUV radiation field ($\chi$) around the gas particle
are calculated at each time step so that the fraction of molecular
hydrogen ($f_{\rm H_2}$) for the gas particle can be derived based on
the ${\rm H_2}$ formation/destruction equilibrium conditions.
Thus the ${\rm H_2}$ fraction for $i$-th gas  particle ($f_{\rm H_2, \it i}$)
is given as;
\begin{equation}
f_{\rm H_2, \it i}=F(T_{\rm g, \it i}, \rho_{\rm H, \it i}, D_i,  \chi_i),
\end{equation}
where $F$ means a function for $f_{\rm H_2, \it i}$ determination.

Since the detail of the derivation methods of $\chi_i$ and $f_{\rm H_2, \it i}$
(thus $F$)  are given
in B13a and B13b, we here briefly describe the methods.
The SEDs of stellar particles around each $i$-th gas particles 
(thus ISRF) are first 
estimated from ages and metallicities of the stars by using stellar population
synthesis codes for a given IMF (e.g., Bruzual \& Charlot 2003).
Then the strength of the FUV-part of the ISRF
is estimated from the SEDs so that $\chi_i$ can be derived for the $i$-th gas particle.
Based on $\chi_i$, $D_i$, and  $\rho_{\rm H, \it i}$ of the gas particle,
we can derive $f_{\rm H_2, \it i}$ (See Figure 1 in B13a).
Thus each gas particle has $f_{\rm H_2, \it i}$, metallicity ([Fe/H]),
and gas density, all of which are used for estimating the IMF slopes
for the particle (when it is converted into a new star).

Since the formation and evolution processes of dust and ${\rm H_2}$ are strongly
coupled in the present study,  the model parameters of dust need to be carefully
chosen so that both the observed ${\rm H_2}$ and dust properties can be well
reproduced by the models. The present model can explain the observed
positive correlation between hydrostatic gas pressure ($P_{\rm gas}$) and
the ratio of ${\rm H_2}$  to H~{\sc i} ($R_{\rm mol}$), 
which can be denoted as $R_{\rm mol} \propto R_{\rm gas}^{0.92}$
(e.g., Blitz et al. 2007). Furthermore, the dust-regulated ${\rm H_2}$ formation
model can explain the observed correlations between local surface densities of
dust and ${\rm H_2}$ in galaxies. The origin of these ${\rm H_2}$ properties
will be discussed in detail by our forthcoming papers in which
the same dust-regulated ${\rm H_2}$ formation model is adopted
(e.g., Bekki 2014).

\subsection{Star formation and SN feedback effects}

Since SF can proceed in molecular clouds,
we adopt the following `${\rm H_2}$-dependent' SF recipe
(B13a) using molecular gas fraction
($f_{\rm H_2}$) defined for each gas particle in the present study.
A gas particle {\it can be} converted
into a new star if (i) the local dynamical time scale is shorter
than the sound crossing time scale (mimicking
the Jeans instability) , (ii) the local velocity
field is identified as being consistent with gravitationally collapsing
(i.e., div {\bf v}$<0$),
and (iii) the local density exceeds a threshold density for star formation ($\rho_{\rm th}$).
We mainly investigate the models with $\rho_{\rm th}=1$ cm$^{-3}$
in the present study.

A gas particle can be regarded as a `SF candidate' gas particle
if the above three SF conditions (i)-(iii) are satisfied.
It could be possible to convert some fraction ($\propto f_{\rm H_2}$)
of a SF candidate  gas particle
into a new star at each time step until the mass of the gas particle
becomes very small. However, this SF conversion method can increase dramatically
the total number of stellar particles, which becomes  numerically very costly.
We therefore adopt the following SF conversion method.
A SF candidate $i$-th gas
particle is regarded as having  a SF probability ($P_{\rm sf}$);
\begin{equation}
P_{\rm sf}=1-\exp ( -C_{\rm eff} f_{\rm H_2}
\Delta t {\rho}^{\alpha_{\rm sf}-1} ),
\end{equation}
where $C_{\rm eff}$ corresponds to a star formation  efficiency (SFE)
in molecular cores and is set to be 1,
$\Delta t$ is the time step width for the gas particle,
$\rho$ is the gas density of the particle,
and $\alpha_{\rm sf}$ is
the power-law slope of the  Kennicutt-Schmidt law
(SFR$\propto \rho_{\rm g}^{\alpha_{\rm sf}}$;  Kennicutt 1998).
A reasonable value of
$\alpha_{\rm sf}=1.5$ is adopted in the present
study. 
This SF probability has been already introduced in our early chemodynamical
simulations of galaxies (e.g., Bekki \& Shioya 1998).

At each time step   random numbers ($R_{\rm sf}$; $0\le R_{\rm sf}  \le 1$)
are generated and compared with $P_{\rm sf}$.
If $R_{\rm sf} < P_{\rm sf}$, then the gas particle can be converted into
a new stellar one.
In this SF recipe, a gas particle with a higher gas density
and thus a shorter SF timescale ($\propto
\rho/\dot{\rho} \propto \rho^{1-\alpha_{\rm sf}}$)
can be more rapidly converted into a new star owing to the larger
$P_{\rm sf}$. Equally, a gas particle with a higher $f_{\rm H_2}$
can be more rapidly converted into a new star.
We thus consider that the present SF model is a good approximation
for star formation in molecular gas of disk galaxies.

Each SN is assumed to eject the feedback energy ($E_{\rm sn}$)
of $10^{51}$ erg and 90\% and 10\% of $E_{\rm sn}$ are used for the increase
of thermal energy (`thermal feedback')
and random motion (`kinematic feedback'), respectively.
The thermal energy is used for the `adiabatic expansion phase', where each SN can remain
adiabatic for a timescale of $t_{\rm adi}$.
Although $t_{\rm adi}=10^5$ yr is reasonable for a single SN explosion,
previous galaxy-scale simulations adopted a much longer timescale
of $\sim 10^7$ yr (e.g., Mori et al. 1999; Stinson et al. 2006),
Multiple  SN explosions can occur for a gas particle with a mass of 
$10^5 {\it M}_{\odot}$ in these galaxy-scale simulations,
and $t_{\rm adi}$ can be different
for multiple SN explosions in a small local region owing to complicated
interaction between gaseous ejecta from different SNe.  
Such interaction of multiple  SN explosions would make the adiabatic phase
significantly longer in real ISM of galaxies.

Considering these, we adopt the following three different 
$t_{\rm adi}$ for modeling SN feedback(SNF): $3\times 10^6$ yr (referred to as 'weak SNF'),
$10^7$ yr ('moderate SNF'), and $3 \times 10^7$ yr ('strong SNF').
We compare the results of these three model with one another to demonstrate
the importance of SNF in the $z$-evolution of dust properties in galaxies.
The energy-ratio of thermal to kinematic feedback is consistent with
previous numerical simulations by Thornton et al. (1998) who investigated
the energy conversion processes of SNe in  detail.
The way to distribute $E_{\rm sn}$ of SNe among neighbor gas particles
is the same as described in B13a.

\subsection{Fixed and time-varying IMFs}

We investigate the dust evolution of galaxies with the fixed Kroupa IMF 
(e.g., Kroupa 2001)
for most models in the present study. However, as shown in B13b,  the time evolution
of dust can be quite different between  fixed and time-varying IMFs. Therefore,
we adopt the time-varying IMF dependent on local gaseous properties (i.e.,
metallicities and  ${\rm H_2}$ densities) and thereby investigate the differences
in dust and ${\rm H_2}$ evolution between the fixed and time-varying IMFs.
The three IMFs slopes ($\alpha_i$, $i=1$, 2, and 3) in the Kroupa IMF
for each gas particle
in the adopted time-varying IMF  depend on the physical properties
of the gas particle as follows:
\begin{equation}
\alpha_i = F_i({\rm [Fe/H]},\rho_{\rm H_2}) 
\end{equation}
where [Fe/H] and $\rho_{\rm H_2}$ are the metallicity and ${\rm H_2}$ density
of the gas particle, and the functional forms are different between different $i$
(i.e., different mass ranges of the IMF).
The adopted IMF form means that the metallicity of a star-forming gas
cloud plays a decisive role in controlling IMF.
The details of the functional form ($F$) and the possible physical
effects of the adopted varying IMF on galaxy evolution are given and 
discussed in B13b.
Although we think that the time-varying IMF is a very important factor for
the $z$-evolution of dust properties in galaxies,
we present the results for just one set of models with the two IMFs.
This is simply because
our main focus is not the roles of IMFs in galaxy formation and evolution in 
the present study.

\subsection{Initial conditions}

The present study is the very first step toward better understand the $z$-evolution
of dust properties in galaxies based on a $\Lambda$CDM cosmology. Therefore, we adopt
idealized initial conditions of galaxy formation based on a $\Lambda$CDM model thereby
determine the reasonable ranges of key parameters for dust growth and destruction
and investigate dust properties of simulated galaxies.
Recent cosmological hydrodynamical simulations of galaxy formation have 
adopted the 'zoom-in' technique to achieve enough spatial resolution to investigate
individual galaxies in a large-scale cosmological initial condition (e.g., the Eris
simulation; Guedes et al. 2011). The present study does not perform such a zoom-in
cosmological simulation, firstly because it is extremely time-consuming to determine
the parameter ranges of dust growth and destruction by running a large number of
zoom-in simulations, and secondly because the adopted GPU-based simulation code
is yet to be well developed for massively parallel GPU clusters essential for such 
zoom-in high-resolution simulations.
Cosmological zoom-in simulations of galaxy formation will be done in our future studies
for a reasonable range of dust parameters derived in the present study.

The way to set up initial conditions for galaxy formation is essentially the same
as that adopted by early numerical studies of galaxy formation based on a CDM
model (Katz 1992; Steinmetz \& M\"uller 1995; Bekki \& Chiba 2000; 
Kawata \&  Gibson 2003). 
We consider an isolated homogeneous, rigidly rotating
sphere on which small-scale initial fluctuations
are superimposed according to a $\Lambda$CDM  power spectrum.  We use the software
GRAFIC (Gaussian Random Fields for Cosmological Simulations;
Bertschinger 1995; 2001) in order to generate initial conditions of galaxy formation.
We adopt a WMAP7 $\Lambda$CDM cosmology with 
($\Omega_{\rm m},\Omega_{\Lambda},\Omega_{\rm b},h,n,\sigma_8$)
=(0.272,0.728,0.046,0.704,0.967,0.810) (Komatsu et al. 2011)  for all simulations. 
The starting redshift ($z_{\rm i}$) for a simulation is determined by GRAFIC for the above
cosmological parameters and the adopted initial mass, size, and particle number
of a galaxy. Typically,  $z_{\rm i}$ is $\sim 16$ in the present study.

The three basic parameters in the present study are the initial total halo
mass ($M_{\rm h}$), spin parameter ($\lambda$), and mass density ($\rho_{\rm h}$).
The spin parameter $\lambda$ is defined as:
\begin{equation}
\lambda= \frac{ J E^{1/2} } { GM_{\rm h}^{5/2} }  ,
\end{equation}
where $J$, $E$, and $G$ are the total angular momentum of a galaxy,  the total
energy of the galaxy, and the gravitational constant. We mainly investigate
the models with $\lambda=0.02$, 0.06, and 0.1. 
In order to discuss the origin of the observed metallicity-dust relation ($A_{\rm O}-D$)
of galaxies, we investigate the models with different $M_{\rm h}$ 
ranging from $3 \times 10^9  {\it M}_{\odot}$ to $10^{12} {\it M}_{\odot}$.
Furthermore, we investigate the models with rather high $\rho_{\rm h}$
so that we can discuss the origin of dust in high-$z$ QSOs.

\subsection{Low- and high-resolution simulations}

One of the main purpose of the present study is to determine the reasonable
range of dust parameters for numerical simulations of galaxy formation. 
The determined dust parameters will be able to be adopted for future more sophisticated
zoom-in simulations of galaxy formation. 
Owing to the limited amount of computing time allocated for the project for
the present GPU-based simulations, we need to take the `two-fold' strategy. 
First, we run a large number of low-resolution simulations ($N \sim 2\times 10^5$)
for different dust
models (and SNF) so that we can determine the reasonable range of dust parameters.
Then, we run high-resolution ($N\sim 10^6$) simulations to investigate the 
dust properties of the simulated galaxies and present the preliminary results.
In this second set of high-resolution
simulations, we try to confirm that the derived dust parameters
can explain observations reasonably well.
The total numbers of dark matter and baryon particles, which are denoted
as $N_{\rm dm}$ and $N_{\rm b}$, respectively, are the same for all simulations
(i.e., $N_{\rm dm}=N_{\rm b}$).

The low-resolution models with $M_{\rm h}=10^{12} {\it M}_{\odot}$ and $\lambda=0.06$
are investigated for different dust parameters so that the final $D$, $q_{\rm PAH}$,
and $f_{\rm H_2}$ can be compared with the corresponding observations for the MW.
The initial radius in a comoving coordinate ($R_{\rm h}$) 
and the total particle number ($N$) are set to be
1.57 Mpc and 282454, respectively (i.e., $\delta x$ (cell size) of
 $0.049$ Mpc in the GRAFIC code),
in the low-resolution models. The values of $\epsilon_{\rm dm}$ and $\epsilon_{\rm g}$
are 1.62 kpc and 0.162 kpc, respectively, and the mass resolution is 
$m_{\rm dm}=5.9 \times 10^6 {\it M}_{\odot}$ for the dark matter halo and
$m_{\rm b}=1.0 \times 10^6 {\it M}_{\odot}$ for the baryonic component (gas).
For the high-resolution models, we mainly investigate the 'fiducial' model corresponding
to the formation of the MW and the parameter values are given in Table 2.
The fiducial model has $N=1041636$, $M_{\rm h}=10^{12} {\it M}_{\odot}$,
$R_{\rm h}=1.59$ Mpc (in a comoving coordinate), 
$\lambda=0.06$,
$\epsilon_{\rm dm}=0.93$ kpc,
$\epsilon_{\rm g}=0.093$ kpc,
$m_{\rm dm}=1.6 \times 10^6 {\it M}_{\odot}$, and
$m_{\rm b}=2.7 \times 10^5 {\it M}_{\odot}$.

The initial masses and sizes  of galaxy formation for high-resolution models
mainly investigated in the present study
are summarized in Table 3. The mass- and size-resolutions are different between
models with different $M_{\rm h}$ and $R_{\rm h}$ for the adopted constant $N=1041636$.
We investigate the high-resolution models with both CDA and VDA and different
SNF (i.e., weak, moderate, and strong). We investigate the roles of time-varying IMFs
in the $z$-evolution of dust and ${\rm H_2}$ properties 
only for the fiducial model in the present study. 
The parameter ranges investigated in the present study are summarized in Table 4.

\subsection{Choice of $\tau_{\rm a}$ and $\tau_{\rm d}$}

As shown in previous one-zone chemical evolution models (e.g., D98)
and numerical simulations (B13a), the most important parameters for 
the time evolution of dust are $\tau_{\rm a}$ and $\tau_{\rm d}$.
Furthermore, these two parameters can control the time evolution
of ${\rm H_2}$ mass fractions ($f_{\rm H_2}$) in galaxies. Therefore,
we need to choose carefully these two parameters in order to reproduce
not only dust properties but also cold gas contents of galaxies in a self-consistent
manner. We therefore first determine the reasonable ranges of these two parameters
by running many low-resolution models with different 
$\tau_{\rm a}$ and $\tau_{\rm d}$. 
Then we try to confirm the selected dust parameters can explain the observed
$D$ and $f_{\rm H_2}$ of the MW
in the high-resolution models.

We compare the observed   $D$, $q_{\rm pah}$, and $f_{\rm H_2}$ 
with those of the Milky-Way (MW)
models in which model parameters for initial conditions of galaxy formation
are chosen for the MW formation.
The time evolution of dust properties can be strongly influenced by
the modeling of SNF, because dust destruction by SNe is included in the present
simulations. We therefore investigate how the dust and ${\rm H_2}$ properties
of galaxies depend on modeling of SNF in the low-resolution MW models too.
For comparison, we run models (i) with dust growth (`W-grow') yet without
dust destruction (`W/O-dest'), (ii) without dust growth (`W/O-grow')
and without dust destruction, and without dust growth and with dust destruction
('W-dest').

Figure 1 shows the comparison between the simulated locations of disk galaxies
at $z=0$ on the $D-f_{\rm H_2}$ and $D-q_{\rm PAH}$ planes with the corresponding
observations for the MW. The observational data points are from Draine \& Li (2006),
Zubko et al. (2004), and Nakanishi et al. (2003, 2006). 
It should be noted here that
there is an uncertainty in the CO-to-${\rm H_2}$
conversion factor, 'X-factor', in estimating $M_{\rm H_2}$ (Nakanishi et al. 2006). 
Both a  constant X-factor of $1.8 \times 10^{20}$ 
(${\rm H_2}$  ${\rm cm}^{-2}$ ${\rm K}^{-1}/$ ${\rm km s}^{-1}$) and
radial-dependent X-factor are adopted for the estimation
of the total ${\rm H_2}$ mass of the Galaxy in Nakanishi et al. (2006).
The observational error bar
of $f_{\rm H_2}$ indicates the possible range of $f_{\rm H_2}$
for a factor of 50\% changes in the X-factor:
The real range of the X-factor difference could be even larger than 50\%.
The two left panes with same dust parameters 
($\tau_{\rm a}=0.25$ Gyr and $\tau_{\rm d}=0.5$ Gyr) yet different
SNF models clearly show that 
the final $D$, $f_{\rm H_2}$, and $q_{\rm PAH}$ are higher for weak SNF.
Star formation is less strongly suppressed in the models with weak SNF
so that chemical evolution and dust enrichment can proceed more efficiently.
Dust is less efficiently destroyed in these models so that ${\rm H_2}$ formation
can proceed more rapidly in these models. As a result of these, 
$f_{\rm H_2}$ can be finally larger in these models.

It appears that the model with moderate SNF can better reproduce the observed
locations of the MW on the two planes. However, as shown in B13a, the evolution
of $D$ and $f_{\rm H_2}$ depends slightly on spatial resolution and minimum
time step width of simulations. Therefore, this result can not be taken as a face value,
and we need to confirm
the dependences of $D$ evolution on SNF modeling later in the high-resolution models.
In the adopted PAH models, the final $q_{\rm PAH}$ appears to be slightly smaller
than the observed one for the three SNF models.  The mass fraction of Carbon metals
from C-rich AGB stars
that can be locked up onto dust grains is assumed to be 0.05 in these models,
because our previous models with this number (0.05) could better explain
the observed $q_{\rm PAH}$ of the MW.
If a larger value of this lock-up fraction is adopted, the final $q_{\rm PAH}$ can be
larger and thus closer to the observation.

The right two panels in Figure 1 shows that if dust destruction is not included,
the final $D$ and $f_{\rm H_2}$ can be too large to be consistent with observations.
Furthermore, if dust growth is not included, then the final $D$ is too small
to be consistent with observations (irrespective of whether dust destruction is included
or not). These three 'non-standard' models also show too small $q_{\rm PAH}$ at
their final $D$. These results clearly demonstrate that both dust growth
and destruction need to be included properly
in numerical simulations of galaxy formation
and evolution for self-consistent reproduction of $D$, $f_{\rm H_2}$, and $q_{\rm PAH}$
in the MW. 

The left two panes in Figure 2 shows that as long as $\tau_{\rm d}=2\tau_{\rm a}$,
the time evolution and the locations of the simulated
galaxies on  the $D-f_{\rm H_2}$ and $M_{\rm dust}-M_{\rm H_2}$ planes
are reasonable for $\tau_{\rm a}=0.125$ Gyr and 0.25 Gyr.
For the model with $\tau_{\rm a}=0.125$ Gyr, $f_{\rm H_2}$
and $M_{\rm H_2}$ at the final $D$ are reasonable too 
even for $\tau_{\rm d}=4\tau_{\rm a}$.
These results for galaxy formation based on a $\Lambda$CDM model
are consistent with those in B13a in which simplified initial conditions
of galaxy formation through gravitational collapse are adopted.
However,  the models with $\tau_{\rm d}=\tau_{\rm a}$
for $\tau_{\rm a}=0.25$ Gyr and 0.5 Gyr
show a very strange behavior in the sense that $f_{\rm H_2}$ and $D$ decrease from some
points owing to the too rapid destruction of dust grains by SNF. 
This behavior of dust and ${\rm H_2}$ content evolution can not be realistic
and such models with $\tau_{\rm d}=\tau_{\rm a}$ should be rejected.
These results demonstrate  that the ratio of dust growth to destruction
timescales is a key for the evolution of dust and ${\rm H_2}$ and thus carefully
chosen in numerical simulations of galaxy formation and evolution. 

The right two panes in Figure 2 shows that even if 
$\tau_{\rm d}=2\tau_{\rm a}$,  the models with rather short
$\tau_{\rm a}$ (=0.01 Gyr and 0.025 Gyr)
can show (i) an unrealistic behavior in the time evolution
of galaxies on the two planes and (ii) the final $f_{\rm H_2}$ inconsistent
with observations. The origin for the rapid decrease of $f_{\rm H_2}$ and
$D$ is that SNF can more rapidly destroy dust grains than SNe and AGB can produce
dust. These results demonstrate that not only the timescale ratio
of dust growth and destruction but also the timescale of dust growth
is important for the time evolution dust and ${\rm H_2}$ properties of galaxies.
It should be noted here that a larger $\tau_{\rm dest}/\tau_{\rm a}$ ($\sim 8$)
is required for the  models with 
$\tau_{\rm a}=0.025$ Gyr to show a realistic behavior of dust and ${\rm H_2}$ content
evolution. 
However, the models with short $\tau_{\rm a}$ yet high $\tau_{\rm d}/\tau_{\rm a}$
show a slightly too large final $\log D$  ($>-1.8$) owing to too rapid growth
of dust. We thus consider that such models with rather short $\tau_{\rm a}$ would not 
be so realistic for dust and ${\rm H_2}$ evolution of galaxies.


\section{Results}

\subsection{Redshift evolution of dust and ${\rm H_2}$ contents}

Figures 3 and 4 show the $z$-evolution of the projected dust mass densities 
distributions  ($\mu_{\rm dust}$) and
the final distributions of stars with different ages and metallicities,
respectively,
for the fiducial model with $M_{\rm h}=10^{12} {\it M}_{\odot}$, $\lambda=0.06$,
the fixed IMF, and
the standard CDA.  Dust densities can become higher in subgalactic clumps at high
$z$ ($3.5 \le z \le 16$),
owing to  the growth of pre-galactic
dust grains (i.e., dust from Pop III stars assumed in this simulation).
Many dusty subgalactic clumps are 
connected with one another by low density filamentary structures at $z=3.5$,
which appears to be a 'cosmic web' of dusty gaseous regions.
Star formation at this stage is not active so that the increase of dust abundances
due to metal ejection from SNe and AGB stars can not occur efficiently.

The interstellar dust of this forming  galaxy can rapidly grow during early
disk formation through  dissipative merging
of dusty gas  clumps ($z=1.8$).  The dust densities
are higher in the inner regions of subgalactic clumps, where star formation
from ${\rm H_2}$ gas clouds is ongoing very efficiently so that dust production rates
from SNe and AGB stars can be very high. 
The merging of these clumps trigger starbursts in the forming galactic gas disk, 
and consequently chemical and dust
enrichment processes can proceed very rapidly.
As a result of this, the dust surface densities in the central regions
of the disk can become significantly higher ($z=1.4$).

The 2D $\mu_{\rm dust}$ distribution shows an intriguing outer ring-like structure,
and the disk has many isolated dusty clumps in its outer halo region at $z=0.4$.
The final galaxy consists of three spatially different components, thin and thick disks,
and old stellar halo at $z=0$,  as already shown in early chemodynamical simulations of 
disk galaxy formation (e.g., Steinmetz \& M\"uller 1995;
Bekki \& Chiba 2000; Brook et al. 2004).
The age-metallicity relation (i.e., higher [Fe/H] for younger ages) can be clearly
seen in this model and the final stellar [Fe/H] at $z=0$ is consistent with
the observed (solar) value. 
Since the origin of these different components of disk galaxies
with different age and metallicity distribution functions  have been discussed
in many simulations already, we do not discuss these in the present study.

Figure 5 shows that although the $z$-evolution
of $M_{\rm g}$ is not so rapid,
the dust-to-gas-ratio ($D$)
can dramatically  increase
during the early disk formation ($z = 1 \sim 3$) via dissipative merging
of gas-rich subgalactic clumps for the fixed  IMF model.  
This strong $D$ increase can be seen
also in the $z$-evolution of  $M_{\rm dust}$
and thus  is  in a striking contrast
with the recent observational result that high-$z$  submillimeter galaxies 
appear to have $M_{\rm dust}$ similar to the optically selected SDSS
galaxies at lower $z$ (D11). 
The slow and steady increase of $D$ at $z>4$ is due simply to the dust growth
via accretion of gas-phase pre-galactic metals onto already existing dust
(i.e., not due to the production of dust from active star formation 
at high $z$).

The total dust mass ($M_{\rm dust}=M_{\rm dust, d}+M_{\rm dust, h}$)
in the models 
with weak SNF and CDA can be significantly
larger than  $10^8 {\it M}_{\odot}$ owing to the less efficient dust
destruction by SN and dust growth in every gas particle. 
As discussed later, the total dust mass of a galaxy at $z=0$ depends
on the modeling of SNF and dust models (CDA or VDA).
As a result of larger $M_{\rm dust}$, the final $D$ in this model is higher 
than the observed $D$ of the MW.
In this high-mass model,  the formation of small dwarf galaxies at very high $z$
can not be well resolved (owing to the adopted particle number)
so that the production of dust from SNe and AGB stars
can be underestimated in the high-$z$ subgalactic clumps.
As a results of this, the formation of ${\rm H_2}$ on dust grains can abruptly
become very efficient at $z\approx 4$
and the time evolution of $M_{\rm H_2}$ and $f_{\rm H_2}$
are very steep around $z=4$. 

The total stellar mass ($M_{\ast}$) increases more sharply
at $z>2$, when both the SFR and sSFR (i.e., SFR/$M_{\ast}$) are rather high.
In the later phase of disk formation ($z<2$), $M_{\ast}$ can grow slowly, which means that
$M_{\rm dust}/M_{\ast}$ does not change so much at lower $z$ 
(owing to the slow change of $M_{\rm dust}$).
The sSFR can be higher at higher $z$, when gas-rich low-mass building blocks
merge with one another to form new stars quite efficiently.
The presence of a peak in the $z$-evolution of the SFR in this model
is in a striking contrast with no peak in that of the sSFR.
It should be stressed here that both $M_{\rm dust}$ and $M_{\ast}$ are the 
total masses of dust and stars, respectively. The total dust mass within 
the disk ($M_{\rm dust,d}$) is significantly smaller than $M_{\rm dust}$
owing to the presence of dusty gaseous halos.

It is intriguing that the $z$-evolution of $f_{\rm H_2}$ 
shows two peaks at $z\sim 2.8$ (earlier) and $z=0.8$ (later).
The first  peak corresponds to the most active star-formation phase
of massive subgalactic clumps, where chemical/dust enrichment processes
and thus ${\rm H_2}$ formation on dust grains are effectively ongoing. 
During  dissipative merging of gas-rich subgalactic clumps at the early disk formation,
${\rm H_2}$ gas can be rapidly consumed by starburst in the gas disk so that
$f_{\rm H_2}$ and $M_{\rm H_2}$ can become slightly smaller.
After the first peak, gradual accretion of gas and the subsequent star formation
can cause dust enrichment and thus effective ${\rm H_2}$ formation on dust grains
so that $f_{\rm H_2}$ can  again start increasing.
Further ${\rm H_2}$ consumption by star formation in the disk can decrease 
$f_{\rm H_2}$ again, which ends up with the second peak.

It is clear that
the $z$-evolution of $M_{\rm g}$, $D$, $M_{\ast}$, and $f_{\rm H_2}$
can be significantly different between the fixed and time-varying IMF models.
During the disk formation with active star formation, the IMF can become
more top-heavy so that a larger amount of metals can be produced and then recycled into
ISM of the forming disk for the time-varying IMF.
 Consequently, a larger amount of dust can be retained in
the disk and a larger amount of ${\rm H_2}$ gas can be processed on dust grains.
The final $D$ and $f_{\rm H_2}$ can be accordingly larger in the time-varying IMF
than in the fixed IMF. These intriguing differences in the time evolution
of $D$ and $f_{\rm H_2}$ have been already reported in B13b for disk galaxy evolution.
It should be noted also that the final $M_{\rm g}$ is smaller
in the time-varying IMF, which means that stronger SNF due to more top-heavy IMF
in active SF regions can suppress the gas consumption by star formation in the model.
The more severe suppression of SF in the time-varying IMF (i.e., slower gas consumption)
can be therefore confirmed in the present study based on the $\Lambda$CDM model.
The present study confirmed that this different evolution 
between the fixed and time-varying IMFs is possible in the
disk galaxy formation based on a  $\Lambda$CDM model.
The basic behaviors of $D$ and $f_{\rm H_2}$ evolution (e.g., two peaks
and rapid evolution in disk formation) are essentially the same between the two
IMF models.

Figure 6 shows that the $z$-evolution of $D$ and
and $q_{\rm PAH}$
depend more strongly on the initial mass densities of dark matter halos than
on the spin parameters ($\lambda$) for a given total mass.
The steep $D$ and $q_{\rm PAH}$ increases come earlier in the model with
the higher initial dark matter density, mainly because star formation and the subsequent
chemical and dust enrichment processes can start in subgalactic clumps earlier
in the model. The observed large $M_{\rm dust}$ and $D$ at a high-$z$ QSO SDSS J1148+5251
($z=6.4$), however, can not be reproduced even by the high-density model in which
$M_{\rm dust}$ and $D$ can start increasing at $z \sim 8$. 
This would mean that the dusty QSO at $z=6.4$ can be formed from very rare high-density
peak and dust growth can be more rapid than the present model predicts. The origin
of this QSO has been discussed by previous works (e.g., Dwek et al. 2007; 
Valiante et at. 2012). We accordingly do not discuss the origin of this object in detail
in the present paper.

Figure 7 confirms that the $z$-evolution of $D$ depends both on the modeling of SNF
and on dust growth and destruction parameters. The final $D$ can be larger 
in the model with weak SNF, mainly because chemical and dust enrichment
can proceed more efficiently owing to  a larger amount of gas  being converted 
into new stars in the model.
The timescale ratio of $\tau_{\rm d}$ to $\tau_{\rm a}$ is important and should be
as large as 2 so that the final $D$ can be within a reasonable range ($\log D \sim -2$)
for the MW-type
disk galaxies.  The models with $\tau_{\rm d}=\tau_{\rm a}$
shows unrealistically small final $\log D$ ($<-3$) at $z=0$, because dust can be
very rapidly destroyed by SNF in these models.
These confirm the results of the low-resolution models shown in Figure 2.

Figure 8 shows that $M_{\rm dust}$ can more sharply increase in more massive
forming disk galaxies  and the final $D$ can be larger in the more massive galaxies.
All of these three disk galaxies with different masses show the two peaks
in $M_{\rm H_2}$ evolution, though the second (later) peak is less clearly seen 
than the first (earlier) one.  
The low mass model ($M_{\rm h}=10^{10} {\it M}_{\odot}$) has a less pronounced
second peak in $f_{\rm H_2}$ evolution,
which is in a striking contrast with other two more
massive models (the first peak in the low mass model is around $z=5.5$,
which is outside the range of the plot in Figure 8).
The first peak in $f_{\rm H_2}$ evolution comes earlier in
forming disk galaxies with lower masses owing to the earlier gravitational
collapse. These results suggest that more massive galaxies can have larger $D$ and
$f_{\rm H_2}$ at $z=0$.

\subsection{$A_{\rm O}-D$ relation}

\subsubsection{Constant Dust accretion model}

Figure 9 shows that the metallicity-dust  ($A_{\rm O}-D$)
relation can be clearly seen in the models with
CDA and weak SNF, though the relation appears to be significantly
shallower than the observation. The simulated range of $A_{\rm O}$ is narrower
than the observed one in spite of the investigated large range of initial
halo masses ($10^{10} \le M_{\rm h}/{\it M}_{\odot}  \le 10^{12}$).
The models with strong SNF show wider ranges of $A_{\rm O}$ and $D$ than
those with weak SNF and thus better reproduce the observed $A_{\rm O}-D$ relation.
However, the simulated slope in the $A_{\rm O}-D$ relation is similar to
that in the models with weak SNF and thus appears to be less consistent 
with observations, though the observation shows a large scatter in D at
a given $A_{\rm O}$.

Furthermore these two models with CDA show significantly higher $D$ at $A_{\rm O}<8$ than
the observed, which means that the present simulations overestimate $D$ in low-mass
galaxies for some reasons.
A possible reason for this will be discussed later in \S 4.
Thus, although the present models with CDA and different SNF
demonstrate that the origin of the 
metallicity-dust relation can be closely associated with the different effectiveness of
SNF in galaxies with different masses,
the observed steeper relation and the large dispersion 
at a given $A_{\rm O}$ are not well reproduced by the models.

\subsubsection{Variable dust accretion model}

CDA dust models assume that the dust growth time scale for each gas particle
depends only on the gas-phase metal abundance. Therefore, dust can grow with
a timescale of dust growth inversely
proportional to the gas-phase metal abundance (i.e., more rapid
growth for higher gas-phase metal abundances),
{\it even if the density of a gas particle is low}. Consequently, $D$ can increase
with time for all particles irrespective of their gas density and temperature,
as long as the particles have  metals.
This way of dust growth in  CDA models
might not be so realistic in real galaxy formation and evolution,
though CDA models were adopted in almost all previous  models of dust
evolution  and successful in reproducing the mean dust properties
of galaxies (e.g., D98; Hirashita 1999; B13a,b).
The dust growth timescale in a gas cloud depends not only on 
the gas-phase metal abundance but also
the temperature and density of gas cloud. Therefore, it is possible that
dust growth can be suppressed in low-density gaseous regions and outer halo gas
in galaxies.

Figure 10 shows the locations of galaxies on  the $A_{\rm O}-D$ plane at $z=16$ (initial),
2, 1, and 0 (present) in  VDA  with strong SNF for
$M_{\rm h}=10^{12} {\it M}_{\odot}$, $10^{10} {\it M}_{\odot}$,
and $3\times 10^9 {\it M}_{\odot}$.  
The final $D$ of less massive galaxies  at $z=0$
becomes lower in these VDA than in CDA so that the 
$A_{\rm O}-D$ relation can be steeper. 
A larger amount of gas (metals and dust)  can be ejected from low-mass galaxies
during galaxy formation 
through SN winds caused by strong SNF.
The ejected gas can have lower gas density so that 
dust growth can not efficiently
occur owing to the assumed dependences of dust growth timescales on
gas density in VDA.
As a result of this, the final $M_{\rm dust}$ and thus $D$ can become significantly
smaller in VDA.

However, the observed  low $\log D$ ($<-4$)
can be better reproduced in the models only if $M_{\rm h}$ is very low 
($3 \times 10^9 {\it M}_{\odot}$). 
The observed very low $\log D$ ($<-4.5$) can not be reproduced by the present models
even for very low initial dark halo masses ($M_{\rm h}=3 \times 10^9 {\it M}_{\odot}$).
It would be intriguing that dwarfs with a wider range of $A_{\rm O}$ ($7.2 < A_{\rm O}<8$)
show such a very low $D$ in observations. 
It would be unreasonable for the present study to explain the observed very low
$D$ by lowering $M_{\rm h}$ to much less than $10^9 {\it M}_{\odot}$.
We thus conclude that the observed very low $D$ could be due to some physical
effects that are not included in the present study.
The cold gaseous components of low-mass forming galaxies can be photo-evaporated  by
ultraviolet background radiation in a high-$z$ reionized universe so that
star-formation (and thus chemical  and dust enrichment processes) can be severely
suppressed (e.g., Susa \& Umemura 2004). This radiative feedback effect is not
included in the present study and thus could be a possible reason for the observed
very low $D$.

Figure 11 shows that the $z$-evolution of dust-to-stellar-mass ratios
($M_{\rm dust}/M_{\ast}$) for $0 \le z \le 0.4$ is rather flat for
the high mass model with $M_{\rm h}=10^{12} {\it M}_{\odot}$.
The observational results by D11 show that
there is a factor of $\sim 5$ difference in $M_{\rm dust}/M_{\ast}$ 
between $z=0$ and 0.4. The models with moderate and strong SNF 
show systematically larger $M_{\rm dust}/M_{\ast}$ and thus the results
are inconsistent with the observed $z$-evolution.
The models with shorter dust destruction timescales ($\beta_{\rm d}=1$ and 1.5)
can show a certain degree of $M_{\rm dust}/M_{\ast}$ decrease,
the models are still unable to reproduce the observed dramatic decrease.
The systematically higher $M_{\rm dust}/M_{\ast}$ can be seen
in the low mass model

It should be stressed here that the total dust mass of a galaxy in the present study
might not correspond to the observed total dust mass, because $M_{\rm dust}$
of a simulated disk includes
both disk and halo dust whereas the observed total
dust of a galaxy would not include halo dust of the galaxy.
We therefore investigated the $z$-evolution of $M_{\rm dust, d}/M_{\ast}$
in the two mass models for a possibly fairer  comparison. 
It is clear in Figure11 that the simulated $M_{\rm dust,d}/M_{\ast}$ does not
show the observed level of dramatic decrease from $z=0.4$ to 0. Since the low-mass
simulated disk has a larger fraction of halo dust (i.e., smaller 
$M_{\rm dust,d}/M_{\rm dust}$), its $M_{\rm dust,d}/M_{\ast}$ can be systematically
lower than the observed value.

The reason of this failure of the present models to reproduce the
observed dust evolution can be  understood as follows.
The total dust masses of galaxies
can decrease mainly through gas consumption due to star formation within the
galaxies
in the present models. 
Gas consumption by star formation between $z=0$ and 0.4 is 
not high so that  dust mass can not decrease so significantly. 
Furthermore, the total stellar mass can only slowly increase
between $z=0$ and 0.4 through the formation of new stars. 
Therefore, the present models fail to reproduce the observed
factor of $\sim 5$ decrease of $M_{\rm dust}$  and $M_{\rm dust}/M_{\ast}$.
We thus conclude that the present models can not reproduce the observed
dramatic evolution of dust masses and dust-to-stellar-mass ratios between
$z=0$ and 0.4.

\subsection{Dust scaling relations at different $z$}

It is useful for the present study to provide some predictions on the dust-scaling
relations, given that 
recent observational studies have started to reveal a number of dust scaling relations of
galaxies (e.g., Cortese et al. 2012).
Four dust and ${\rm H_2}$ scaling relations are derived for 18 models with VDA 
and the results are described in Figure. 12-15.
Figure 12 shows that a clear positive correlation between $M_{\ast}$ (total stellar mass)
and $M_{\rm dust}$ 
(i.e., larger $M_{\rm dust}$ for larger $M_{\ast}$)
is already established at $z=2$ in the simulated galaxies. 
The simulated correlation can be fit roughly to $M_{\rm dust} \propto M_{\ast}^{1.1}$
and does not appear to change significantly between $z=2$ and $z=0$. The fact that
the $M_{\ast}-M_{\rm dust}$ relation is less clear at $z=5$ suggests that the correlation
can be established during the early formation of galactic disks ($z = 1\sim 2$). 
The correlation does not show a large dispersion at a given $M_{\ast}$ in the 
present simulations.

As shown in Figure 13,  the simulated galaxies with larger $M_{\rm dust}$ have larger
$M_{\rm H_2}$ at $z=0$, 1, and 2, though the positive correlation can not be clearly
seen at $z=5$. The simulated $M_{\rm dust}-M_{\rm H_2}$ is closer to the observed
one by Corberi et al. (C12) at $z=2$ and starts to deviate from the observed
one for galaxies with lower $M_{\rm dust}$ ($<10^8 {\it M}_{\odot}$) at $z=1$.
The differences in $M_{\rm H_2}$ between the observed and simulated correlations
become very clear for galaxies with lower dust masses at $z=0$. 
The simulated lower $M_{\rm H_2}$ results from the fact that  low-mass
and low-metallicity galaxies can experience both the steady consumption
of $M_{\rm H_2}$ and much less efficient ${\rm H_2}$ production 
(owing to low metallicity/dust contents and low gas densities) at later $z$.
This inconsistency between the observed and simulated $M_{\rm dust}-M_{\rm H_2}$
correlations for low $M_{\rm dust}$ implies that
some physical effects missing in the present study
would need to be included in our future work.

Figure 14 shows that the simulated galaxies with larger $M_{\ast}$
show larger $f_{\rm H_2}$ 
at $z=0$, 1, and 2, though such a clear positive correlation can not be seen at $z=5$.
As shown in Figure 12,  galaxies with larger $M_{\ast}$ can have larger $M_{\rm dust}$
so that the formation of ${\rm H_2}$ gas can be more efficient in the galaxies.
Therefore, it is reasonable that galaxies with larger $M_{\ast}$ have larger
$f_{\rm H_2}$.  It seems that (i) the dispersion in $f_{\rm H_2}$ for a given
$M_{\ast}$ becomes smaller 
at lower $z$ in the present simulations
and (ii) $f_{\rm H_2}$ for a given $M_{\ast}$ is higher at higher $z$.
These results will be able to be compared with future ALMA observations of
total ${\rm H_2}$ gas mass in galaxies at different $z$.

As shown in Figure 15,  the simulated galaxies have  a positive correlation
between $f_{\rm H_2}$ and $M_{\rm dust}$ at $z=0$, 1, and 2, though the slope
of the $M_{\rm dust}-f_{\rm H_2}$ correlation appears to be slightly different
at different $z$. The simulated almost linear (i.e., steep) correlation implies
that low-mass galaxies with low $M_{\rm dust}$ can form stars much less productively
owing to very low ${\rm H_2}$ masses. The predicted tight $M_{\rm dust}-f_{\rm H_2}$
correlation at $z=0$ (and lower $z$)  can be tested against ongoing 
and future observations for a wide range of $M_{\rm dust}$.

\subsection{Radial gradients of $D$ and dust depletion levels}

Figure 16 shows that the simulated disk galaxy at $z=1.4$ in the MW model
with VDA has a negative radial gradient of $D$ (i.e., larger $D$ in smaller $R$),
though the dispersion of $D$ at a given $R$ is quite large at this epoch.
The main reason for this negative radial gradient is that the disk can have
rather small $D$ in the outer region where chemical enrichment has not yet well proceeded
at $z=1.4$ when the disk is still forming via dissipative merging.
It should be stressed here that although different local regions (marked by small red dots)
show quite different $D$, [S/Fe] (corresponding to dust-depletion level),
and [S/H] (corresponding to gas-phase
metallicity that is not influenced by dust-depletion), both $D$ and [S/H]
show negative radial gradients. 
The derived  negative $D$ and $A_{\rm O}$ gradients
in the present models,
which show large $f_{\rm dust}$ variation, 
are consistent with the recent observations for M31 (e.g., Draine et al. 2014).

The radial gradient of $D$ later becomes shallower at $z=0$ owing to chemical
and dust enrichment processes in the outer part of the gas disk.
Gas-phase chemical abundances ([Fe/H] and [S/H]) in the disk show  slightly negative 
radial gradients and their dispersions are quite large at $z=1.4$.
The gradients become almost flat with smaller dispersions in the gas disk at $z=0$.
Thus, the radial gradients of $D$ and gas-phase chemical abundances 
evolve significantly with $z$
for the MW model with VDA. 

In the present dust models,  gas-phase [S/Fe] can be used as a
measurement of dust depletion for each gas particle,
because dust depletion is not included for S
whereas it is properly included for Fe. 
The radial gradient of [S/Fe] at $z=1.4$ is negative (i.e., larger in smaller $R$)
and steeper  with 
increasingly larger dispersions at larger $R$ in this model. The derived dispersion
in [S/Fe] means that dust depletion levels can be quite different between different
individual
local gaseous regions in a forming galaxy. The radial gradient of [S/Fe] at $z=0$ is
flatter
with rather small dispersions at different $R$, which suggests that the spatial
variations of dust depletion
levels can be more clearly seen in the high-$z$ galaxies. 
These derived spatial and temporal variations of dust depletion levels 
have some important implications on observational and theoretical studies of
galaxy formation, which are later discussed.

As shown in Figure 17, the radial gradients of $D$ at $z=0$ can be quite different
in galaxies with different ${\it M}_{\rm h}$, $\lambda$, and $\rho_{\rm h}$ for 
a given set of dust parameters.  Although all of the models in Figure 17 show the negative
radial gradients of $D$, the gradients are less clear for low-mass models with
$M_{\rm h}=10^{10} {\it M}_{\odot}$. The slopes of the
radial $D$ gradients are significantly
shallower than $-0.04$ (dex kpc$^{-1}$) that is observed for the metallicity gradient
of the Galaxy (e.g., Andrievsky et al. 2004). 
The negative radial gradients of $D$ less steeper than gas-phase metallicity gradients
can be tested against ongoing observations.

Figure 18 shows that the dust-to-metal ratios ($f_{\rm dust}$) are quite different 
between different individual gaseous regions of a single galaxy at $z=0$, 1.4, and 1.8
and $f_{\rm dust}$ can be quite different for a given $D$.  The mean $f_{\rm dust}$
evolves with $z$ such that it can become larger (i.e., more strongly dust-depleted) 
at lower $z$. The derived temporal and spatial variations of $f_{\rm dust}$ strongly
suggest that a fixed dust-to-metal ratios should not be adopted in 
theoretical works on the SEDs of galaxies and the formation and evolution
of cold ${\rm H_2}$ gas. The dispersion of $f_{\rm H_2}$ at a given $D$ can be clearly
seen at the three $z$ in this model, which demonstrates
 that larger $D$ does not necessarily
mean higher efficiencies of ${\rm H_2}$ formation in galaxies.

The main physical reason for large $f_{\rm dust}$ variation is that different
local regions experience different chemical evolution histories and dust accretion
and destruction processes caused by different physical conditions of ISM and
star formation histories. Therefore, the derived $f_{\rm dust}$ variation
is not due to the adopted models  (e.g., star formation dependent on ${\rm H_2}$ 
densities) in the present simulations. Such a large temporal variation
is already found in previous one zone models, e.g., Figure 6 in Inoue (2003)
and Figure 7 in Calura et al. (2008).

Recent observational studies on dust-to-metal
ratios in damped Lyman-$\alpha$ absorbers (DLAs) at higher $z$ ($>1.2$)
have revealed a large (a factor of $\sim 10$)
variation
of $f_{\rm dust}$ in the ISM of the DLAs (e.g., De Cia et al. 2013).
Furthermore, $D$ at a given $A_{\rm O}$ in the observed $D-A_{\rm O}$
relation shows rather large dispersion
(G11), which suggests that dust-to-metal-ratio is not constant among
different galaxies.
These observations are consistent with the large $f_{\rm dust}$ variation
derived in the present study.
The possible temporal and spatial  $f_{\rm dust}$ variations
would have some implications on 
some recent observational works, e.g.,
Eales et al. (2012),  which have tried to estimate the total amount of cold gas in galaxies
by using the continuum dust emission,
and Grootes et al. (2013), which have revealed correlations between
the stellar mass surface densities and the optical depth due to dust in galaxies.

\subsection{Formation of extended dusty gaseous halos}

The simulated negative radial gradient of $D$ in Figure 16 implies that the projected
radial density profile of dust in a galaxy can have a steep gradient. Figure 19
shows that the simulated disk galaxies can have gaseous halos ($R>30$ kpc)
extended well beyond the disk sizes ($\sim 20$ kpc).
The radial profile
of the cumulative dust mass ($M(<R)$) normalized by the total mass within 100kpc 
($M(R<100$kpc)) clearly demonstrate that a larger fraction of dust in the outer
($R>30$) halo of the simulated disk galaxies. The extended dusty gaseous halos
can be formed in most of the present models.
Since the present chemodynamical models do not include the dust destruction
by hot gaseous plasma in galaxy halos, they might have overestimated the
total amount of halo dust.

The simulated slope in the mass profiles of dust in the
outer halo ($R>30$  kpc) are too steep to be consistent
with the observed one ($\Sigma \propto R^{-0.8}$) by 
M\'enard et al 2010 (M10). It is currently unclear
why the observed slope is significantly shallower than the mass distributions
of gas and dark matter halos in galaxies.
Possibly, we would need to incorporate
an additional  physical process for dust evolution (e.g., radiation pressure
on  dust grains) into chemodynamical simulations 
in order to reproduce the observed shallower profile of dust.

During the hierarchical build-up
of the disk galaxies,  their cold gas containing dust can be expelled from the galaxies
owing to the energetic 
SNF. The expelled dust can not be return back to the host galaxies
quickly so that they can be observed as being in the halo regions. 
This `blown-off'  effect of SNe is the key physical process for the formation
of dusty gaseous halos in galaxies.
As shown in Figure 11, the mass fractions of halo dust ($M_{\rm dust, h}/M_{\rm dust}$)
can be  larger in less massive galaxies. This is mainly because
the less massive galaxies are more prone to the energetic SNF during their formation.
It is observationally unclear, however, whether less massive disk galaxies
have larger fractions of dust in their halos.

\section{Discussion}

\subsection{Origin of the observed $A_{\rm O}-D$  relation}

Recent observations have found the following three characteristics
of  the metallicity-dust relation
in galaxies (e.g., G11).
First, more metal-rich galaxies with larger $A_{\rm O}$ can show
larger $D$. 
Second, the dispersion in $D$ at a given $A_{\rm O}$ is much larger than the possible
observational error in the estimation of $D$.
Third, the metallicity-dust relation appears to be different for $A_{\rm O} <8$
and $A_{\rm O} \ge 8$). The linear $A_{\rm O}-D$ correlation for $A_{\rm O}<8$
appears to be steeper than that for $A_{\rm O} \ge 8$.
The present study has shown that more massive galaxies can show both larger $A_{\rm O}$
and larger $D$, which is qualitatively consistent with the observed relation.
Furthermore, the simulated relation is steeper and thus more consistent
with the observed one, if the strong SNF model 
is adopted. 
The main physical reason for the larger $D$ in more massive galaxies
is that a larger fraction of dust and metals can be retained 
in the galaxies (i.e., removal of dust and metals through energetic stellar winds is
not so effective in more massive galaxies).

Although the present study has demonstrated that
SNF is a key physical process for the formation of the metallicity-dust relation,
the observed large scatter for a given $A_{\rm O}$ can not be well
reproduced in the present study.  One possible reason for this is that
the present simulations do not include dust removal through radiation pressure of
stars on dust grains. If only dust (not gas-phase metals) can be selectively removed
by this radiation process, then $D$ can be even lower for a given $A_{\rm O}$.
This removal of dust through dust winds caused by stellar radiation pressure
might depend on
a number of physical parameters of dust and galaxies, such as size distributions of
dust grains and age distributions of stars in galaxies. Therefore, the dust removal
efficiency might be different between galaxies with a similar $A_{\rm O}$ so that
galaxies with a similar $A_{\rm O}$ can show a larger dispersion in $D$.
In order to discuss whether the selective dust removal through stellar radiation
is possible,
we will include the dust removal process 
in our future  more sophisticated simulations.

The observed apparently steeper slope for $A_{\rm O}<8$ can not be clearly seen
in the present study. The observed presence of dwarfs with $A_{\rm O}<8$ and
$\log D <-4.5$ at $z=0$ can not be well reproduced by the present study either.
The observed very low $D$ in dwarfs implies that dust is either destroyed 
or removed more
efficiently than the present simulations predict. Some of the observed dwarfs at $z=0$
show a very low $\log D$ ($<-5$), which is very hard to be explained by the present study.
This inconsistency between observations and simulations clearly suggests that
some physical effects related to dust formation and evolution missing in
the present simulations need to be included in our future simulations.

\subsection{Temporal and spatial variations of  dust-to-metal-ratios in galaxies}

The present study has shown that dust-to-metal ratios ($f_{\rm dust}$)
of galaxies evolve with $z$
and they can be quite different between different local gaseous regions in an individual
galaxy at a given $z$. This result provides the following implications
on the observational and theoretical derivation of galactic SEDs and
the construction of ${\rm H_2}$ formation models in numerical studies of galaxy
formation and evolution.
Numerical simulations of galaxy formation and evolution tried to predict the SEDs
of dusty galaxies based on the age and metallicity distributions of stars and the spatial
distributions of gas and stars in simulated galaxies
(e.g., Bekki et al. 1999; Bekki \& Shioya 2000, 2001; Jonsson 2006).
A basic assumption in these previous works is that $f_{\rm dust}$ is fixed
at a certain value (e.g., an observed value for the Galaxy)
for all gas particles at a given time step. The present results suggest
that such an assumption of constant $f_{\rm dust}$ is not reasonable,
though it is unclear how much the previous derivation of galactic SEDs
can change if $f_{\rm dust}$ variation is correctly included in their simulations.
Future numerical simulations of dusty galaxies will need to include
spatial and temporal variations of $f_{\rm dust}$  in order to predict the SEDs
more precisely.

Numerical simulations of galaxy formation and evolution
and theoretical models of star formation  have just recently started 
to include ${\rm H_2}$ formation on dust grains in ISM 
(e.g.,  Pelupessy et al. 2006; Krumholz, McKee \& Tumlinson 2009; Christenson et al. 2012).
Although the formation of ${\rm H_2}$ on dust grains depend on dust abundances,
compositions, and size-distributions,  the formation efficiencies of ${\rm H_2}$ on
dust grains in these works is assumed to be a function of metallicity (i.e.,
dust abundances is simply proportional to metallicity). The present results
mean that the assumed constant $f_{\rm dust}$
is clearly an over-simplification for the ${\rm H_2}$ formation 
processes and thus that the previous simulations might have over- or under-estimated
the total ${\rm H_2}$ masses of galaxies. 
Future numerical simulations of galaxy formation and evolution with ${\rm H_2}$ formation
will need to include the time evolution of $f_{\rm dust}$ in order to
predict ${\rm H_2}$ contents of galaxies more precisely.

\subsection{Consistency with observations}

Although the present study has shown only the preliminary results of the simulations,
it would be meaningful for the study to discuss the consistency between the simulated
dust properties of galaxies and the corresponding observational ones. 
If the observed properties are not reproduced by the present models,
then additional physical effects on dust grains that are not included in the 
models should be considered in future more sophisticated models.
Table 5 lists up 13 physical properties of galaxies for which observations
and simulations are compared with. 

Only three observations among thirteen can be reproduced reasonably well by the 
present model, which means either  that some key physical processes related to
dust formation and evolution are not incorporated properly in the model
or that the resolution of numerical simulations or adopted idealized initial
conditions can be responsible for the  less successful reproduction.
Clearly, the influences of stellar radiation on dust grains 
(e.g., Ferrara et al. 1991; Bekki \& Tsujimoto 2014)
is completely ignored, though it can remove the dust from gas disks of galaxies.
This ignorance might lead us to  overestimate the dust production rate within galaxies.
Furthermore, the dust-gas  interaction, 
which is demonstrated to be important in galactic dynamics for large dust mass
fractions (Theis \& Orlova 2004), is also ignored in the present study.
We will need to include these missing ingredients in our future simulations to address 
the latest observations of galactic  dust evolution in detail.

Lastly, we briefly discuss the present dust-regulated star formation model by comparing
the observed correlations of sSFRs with  $z$  and $M_{\ast}$ (Elbaz et al. 2011, E11)
with the corresponding simulation results. Figure 20 shows that the sSFRs are systematically
lower than those for galaxies on the `main sequence' at lower $z$ ($<2$). 
Furthermore, sSFRs of the two simulated disk galaxies
at $z=0$ seem to be lower for their large $M_{\ast}$.
These results imply that the present dust-regulated SF model (or ${\rm H_2}$-dependent
SF model) might underestimate the star formation efficiency in ISM of galaxies.
We need to discuss this point in a separate paper.

\section{Conclusions}

We have investigated the $z$-evolution of dust and gas contents
in galaxies and its dependences on initial conditions
of galaxy formation models based on a $\Lambda$CDM cosmology
by using our chemodynamical simulations with dust growth
and destruction.
This is the very first application of our new chemodynamical simulation
code with dust evolution to a numerical study of hierarchical galaxy formation.
We therefore needed to  derive  a reasonable range of dust parameters
that explain observations by running a large number of
models based on  somewhat idealized initial conditions.
The derived
dust parameters will be used for future more sophisticated zoom-in galaxy formation
simulations.

In this preliminary investigation,
we have focused particularly on the $z$-evolution of 
total dust and ${\rm H_2}$ masses ($M_{\rm dust}$ and $M_{\rm H_2}$, respectively), 
dust-to-gas-ratios
($D$), molecular hydrogen fraction ($f_{\rm H_2}$),
dust depletion levels ([S/Fe] and $f_{\rm dust}$),
gas-phase abundances ($A_{\rm O}\equiv$12+log(O/H)),
and radial gradients of $D$ and gas-phase abundances.
We have also investigated
correlations and scaling-relations  between these properties and
the dependences of dust and ${\rm H}_2$ properties on
initial mass densities, spin parameters, and masses of galaxies.
The present preliminary results are as follows \\

(1) The dust growth and destruction timescales 
($\tau_{\rm a}$ and $\tau_{\rm d}$, respectively) need to be carefully
modeled so that both the present dust and ${\rm H_2}$ properties of luminous
galaxies can be reproduced reasonably well. 
For example, $\tau_{\rm a}$ as long as $\sim 0.1$ Gyr is required for the models
with CDA to explain the observed dust and ${\rm H_2}$ properties of the Galaxy.
Without dust growth and destruction,
the observed $D$ and $f_{\rm H_2}$ can not be well reproduced by the present simulations.
These results imply that dust evolution needs to be  properly included 
in numerical simulations of galaxy formation for predicting not only
dust properties of the present galaxies
but also the $z$-evolution of ${\rm H_2}$ contents in galaxies.

(2) Both $D$ and $q_{\rm PAH}$ can grow rapidly during the early dissipative
formation of galactic disks through merging of gas-rich subgalactic clumps 
at $z \approx 2-3$. The $z$-evolution of $D$ and $q_{\rm PAH}$ in
disk galaxies depends strongly on
mass densities of forming galaxies in such a way that
$D$ and $q_{\rm PAH}$ evolution can start increasing earlier and  
the commencement of the rapid $q_{\rm PAH}$ increase is slightly later than that of the
rapid $D$ increase in galaxies. The time evolution of $D$ and $q_{\rm PAH}$ does
not depend so strongly on spin parameters ($\lambda$) for a given initial mass.
 
(3) Disk galaxies with lower initial masses can lose a larger amount of 
dust  through SN winds so that
their final $D$ and $A_{\rm O}$ can be smaller.
The present variable dust growth model (VDA) can better reproduce
the observed $A_{\rm O}-D$ relation than the fixed dust growth model (CDA). However
the observed relation shows a significantly larger dispersion of $D$ at a given
$A_{\rm O}$ than the simulated one.
The present simulations do not reproduce the observed dwarfs
with low $A_{\rm O}$ ($<8$) and very low $\log D$ ($<-4.5$) at $z=0$.
This failure of the present models to reproduce such a low $D$ at $z=0$
could be related to the fact that the present models do not include dust removal
through dust wind due to radiation pressure of stars.

(4) The simulated disk galaxies show little evolution of $M_{\rm dust}$
($M_{\rm dust, d}$) and $M_{\rm dust}/M_{\ast}$ ($M_{\rm dust,d}/M_{\ast}$) 
between $z=0$ and 0.4. The observed dramatic evolution (a factor of $\sim 5$ differences
between $z=0$ and 0.4) of these dust properties
derived by D11 is thus inconsistent with the present
results.
The total amount of ISM with dust can be decreased by gas consumption 
due to  star formation
in the simulated disk galaxies, but the rate of gas/dust consumption
can not be rapid 
enough to explain the observed $z$-evolution of dust properties.
These imply that some key physics related to dust evolution
is missing in the present chemodynamical simulations.

(5) Disk galaxies at $z=0$ show negative radial gradients
of $D$ (larger in smaller radii from galaxy centers) for most models,
and the 
gradients evolve with $z$ in luminous disk galaxies
(i.e., steeper at higher $z$). Both $D$ and gas-phase [Fe/H] are significantly
different between different local regions in an individual galaxy
owing to different chemical
evolution and SF histories in these regions.
Dwarfs (or low-mass disk galaxies) can have dusty gaseous halo that is formed
through ejection of dust due to supernova winds.

(6) Dust depletion levels estimated from gas-phase [S/Fe] 
and $f_{\rm dust}$
are spatially different between individual local gaseous
regions of  a single galaxy and between different galaxies
at a given $z$.  The dust depletion levels  evolve  with $z$
owing to chemical and dust evolution of galaxies.
These results imply that dust-to-metal-ratios should not be fixed in constructing
SEDs of galaxies and predicting ${\rm H_2}$ evolution  of galaxies.

(7) The simulated disk galaxies can have very extended dusty gaseous halos with the
mass fractions ($M_{\rm dust, h}/M_{\rm dust}$) as large as or larger than 0.5.
This could be overestimation, because dust destruction by hot gaseous halo
is not included in the present simulations.
During the formation of disk galaxies,   the cold gas with dust can be blown off
by energetic SNF.  Since less massive galaxies are more prone to such SNF,
the mass fractions of halo dust are higher. The projected radial density
profiles of the dusty halos are rather steep and thus inconsistent with
the observed ones ($\Sigma \propto R^{-0.8}$).

(8) The simulated $M_{\rm dust}-M_{\rm H_2}$ scaling relation is steeper than
the observe one for low-mass galaxies with lower $M_{\rm dust}$. The present simulations
predict that galaxies with larger $M_{\ast}$ have larger $M_{\rm dust}$ 
($M_{\rm dust} \propto M_{\ast}$) and 
$f_{\rm H_2}$ ($f_{\rm H_2} \propto M_{\ast}$) . These scaling relations can be 
seen already at $z=2$, though they are less clear at $z=5$.
There is a strong correlation between $f_{\rm H_2}$ and $M_{\rm dust}$
($f_{\rm H_2} \propto M_{\rm dust}$) in simulated galaxies at $z=0$.

(9) The final $D$ and $f_{\rm H_2}$ in disk galaxies at $z=0$
are larger in time-varying  initial mass functions
(IMFs) than in fixed ones, mainly because
chemical evolution can proceed more rapidly in the time-varying IMFs.
Furthermore, gas consumption rates (i.e., SFRs) and final total stellar masses
are lower and smaller, respectively, in the models with the time-varying IMFs.
Since the present study did not investigate the models with time-varying IMFs in detail,
our future works need to confirm this importance of IMFs in the $z$-evolution
of dust and ${\rm H_2}$ properties of galaxies.

(10) Given a number of inconsistencies between the observed and simulated dust
properties of galaxies in the present study, 
we will need to improve the dust models and the way to incorporate the dust
physics into the chemodynamical models in our future works.
In particular, the models for dust accretion and destruction would need to be
more sophisticated by including the time evolution of dust sizes and
the locally different destruction processes of dust by SNe.
The influences of stellar radiation on dust grains and dust-gas interaction
in ISM are among the new ingredients that should be also included in our future
more sophisticated numerical simulations of galaxy formation.

\section{Acknowledgment} 
I (Kenji Bekki; KB) am   grateful to the referee  for  constructive and
useful comments that improved this paper.
Numerical simulations  reported here were carried out on
the three GPU clusters,  Pleiades, Fornax,
and gSTAR kindly made available by International Center for radio astronomy research
(ICRAR) at  The University of Western Australia,
iVEC,  and the Center for Astrophysics and Supercomputing
in the Swinburne University, respectively.
This research was supported by resources awarded under the Astronomy Australia Ltd's ASTAC scheme on Swinburne with support from the Australian government. gSTAR is funded by Swinburne and the Australian Government's
Education Investment Fund.
KB is grateful to Loretta Dunne for her providing the observational data
sets for dust-to-stellar-mass ratios from her observations.
KB is also grateful to Cameron Yozin-Smith for his constructing a table for observational
data for dust-to-gas-ratios.
KB acknowledges the financial support of the Australian Research Council
throughout the course of this work.


\begin{deluxetable}{ll}
\footnotesize  
\tablecaption{ Description of physical meanings for 
symbols often used in the present study.
\label{tbl-1}}
\tablewidth{-2pt}
\tablehead{
\colhead{  Symbol } &
\colhead{  Physical meaning } }
\startdata
CDA & constant dust accretion (model)   \\
VDA & variable dust accretion (model)   \\
SNF & supernova feedback (effect)   \\
$\tau_{\rm a}$ & dust accretion timescale   \\
$\tau_{\rm d}$ & dust destruction  timescale   \\
$\tau_{\rm a,0}$ & a constant for dust accretion timescale (VDA)  \\
$\beta_{\rm d}$ & $\tau_{\rm d}/\tau_{\rm a}$   \\
$A_{\rm O}$  & gas-phase oxygen abundances  \\
$D$   &  dust-to-gas-ratio \\
$f_{\rm H_2}$ & mass fraction of molecular hydrogen (${\rm H_2}$) \\
$f_{\rm dust}$ & dust-to-metal-ratio \\
$q_{\rm PAH}$ & PAH-to-dust-ratio \\
$R_{\rm 0.5}$ & half-mass radius of stars   \\
$M_{\rm h}$ & initial  halo mass   \\
$M_{\ast}$ & total stellar  mass   \\
$M_{\rm g}$ & total gas  mass   \\
$M_{\rm dust}$ & total dust  mass   \\
$M_{\rm dust,d}$ & total dust  mass in disk ($R\le R_{\rm 0.5}$) \\
$M_{\rm dust,h}$ & total dust  mass in halo  ($R > R_{\rm 0.5}$) \\
$M_{\rm PAH}$ & total PAH dust   mass   \\
$M_{\rm H_2}$ & total ${\rm H_2}$  mass   \\
$\mu_{\rm dust}$ &  surface dust mass density    \\
\enddata
\end{deluxetable}

\begin{deluxetable}{ll}
\footnotesize  
\tablecaption{
Description of the basic parameter values
for the fiducial model.
\label{tbl-2}}
\tablewidth{-2pt}
\tablehead{
\colhead{ Physical properties }
& colhead{ Parameter values } }
\startdata
Cosmological parameters & WMAP 7yr \\
{Total Mass 
\tablenotemark{a} }
& $M_{\rm h}=10^{12} {\it M}_{\odot}$  \\
{Initial size  
\tablenotemark{b} }
& $R_{\rm h}=1.59$ Mpc  \\
Gas fraction & $f_{\rm g}=0.17$     \\
Spin parameter &   $\lambda=0.06$  \\
Feedback & Supernova feedback (SNF) only \\
{SNF strength  
\tablenotemark{c} }
& weak \\
{SNIa model 
\tablenotemark{d} }
& Prompt, $f_{\rm b}=0.05$ \\
Chemical evolution & Non-instantaneous recycling \\
Chemical yield  &  T95 for SN,  VG97 for AGB \\
Initial metallicity   &   ${\rm [Fe/H]_0}=-3$ \\
Dust accretion & Constant growth (CDA) \\
Selective dust depletion & Yes \\
Dust timescales  & $\tau_{\rm acc}=0.25$ Gyr, $\tau_{\rm dest}=0.5$ Gyr  \\
{PAH 
\tablenotemark{e} }
&  $R_{\rm PAH}=0.05$ \\
Dust yield  &  B13a  \\
Initial dust/metal ratio  & 0.1  \\
{SF 
\tablenotemark{f} }
& ${\rm H}_2$-dependent,  ISRF,  $\rho_{\rm th}=1$ cm$^{-3}$ \\
{IMF 
\tablenotemark{g} }
& Fixed Kroupa  \\
Softening length  & $\epsilon_{\rm dm}=0.93$ kpc, 
$\epsilon_{\rm g}=0.093$ kpc \\
Gas mass resolution   & $m_{\rm g}=2.7 \times 10^5 {\it M}_{\odot}$ \\
\enddata
\tablenotetext{a}{
$M_{\rm h}=M_{\rm dm}+M_{\rm g}$, where
$M_{\rm dm}$ and $M_{\rm g}$ are the total masses of dark matter halo
and gas in a galaxy, respectively. The gas fraction ($f_{\rm g}$)
is $M_{\rm g}/M_{\rm dm}$.
}
\tablenotetext{b}{
The size of a gas sphere in a comoving coordinate.
}
\tablenotetext{c}{
Weak SNF' means that $t_{\rm adi}=3 \times 10^6$ yr,
where $t_{\rm adi}$ is the adiabatic phase of (multiple) supernova explosion.
}
\tablenotetext{d}{
$f_{\rm b}$ is the binary fraction of stars that
can finally explode as SNe Ia.
}
\tablenotetext{e}{
$R_{\rm PAH}$ is the
mass fraction of PAH dust to total dust in the stellar ejecta
of C-rich AGB stars.
}
\tablenotetext{f}{
$\rho_{\rm th}$ is the threshold gas density for star formation
and interstellar radiation field (ISRF) is included in the estimation of
${\rm H_2}$ mass fraction in this model.
}
\tablenotetext{g}{
The time-varying Kroupa IMF (B13b) model  is also investigated
just for comparison.
}
\end{deluxetable}

\begin{deluxetable}{ccc}
\footnotesize  
\tablecaption{
Basic parameters for main  models with different initial masses and sizes.
\label{tbl-1}}
\tablewidth{-2pt}
\tablehead{
\colhead{ Model/Parameters  \tablenotemark{a} } & 
\colhead{ $M_{\rm h}$ (${\it M}_{\odot}$)  \tablenotemark{b} } &
\colhead{ $R_{\rm h}$ (Mpc) \tablenotemark{c} } } 
\startdata
M1 &  $10^{12}$ & 1.59 \\
M2 &  $3 \times  10^{11}$ & 1.01 \\
M3 &  $10^{11}$ & 0.91 \\
M4 &  $3 \times 10^{10}$ & 0.46 \\
M5 &  $10^{10}$ & 0.30 \\
M6 &  $3 \times 10^{9}$ & 0.20 \\
\enddata
\tablenotetext{a}{
The model M1 corresponds to the fiducial model in which
the dust and ${\rm H_2}$ properties at $z=0$ can be compared with those of the MW.
Accordingly, the model is also referred to as the MW model. The initial mass density
of this model is defined as $\rho_{\rm h, f}$. The low-density model
with $\rho_{\rm h}=0.5\rho_{\rm h, f}$ and the high-density model
with $\rho_{\rm h}=2\rho_{\rm h,f}$ are also investigated.
}
\tablenotetext{b}{
The initial total mass of a galaxy.
}
\tablenotetext{c}{
The initial size of a gas sphere in a comoving coordinate.
}
\end{deluxetable}

\begin{deluxetable}{ccccccc}
\footnotesize  
\tablecaption{
A range of model parameters.
\label{tbl-4}}
\tablewidth{-2pt}
\tablehead{
\colhead{ Parameters } &
\colhead{ $M_{\rm h}$ (${\it M}_{\odot}$) } &
\colhead{  $R_{\rm h}$ (Mpc) }  &
\colhead{ $\lambda$ \tablenotemark{a} } &
\colhead{ Dust growth} &
\colhead{ $\tau_{\rm a}$ (Gyr) \tablenotemark{b} } &
\colhead{ $\beta_{\rm d}$ \tablenotemark{c} }  }
\startdata
Range &
$3 \times 10^{9}-10^{12}$
& $0.2-1.99$
& $0.02-0.1$
& CDA or VDA
& $0.01-0.74$
& $1-8$
\enddata
\tablenotetext{a}{
We mainly investigate the models
with $\lambda=0.06$, though other models with  $\lambda=0.02$ and 
0.1 are investigated.
}
\tablenotetext{b}{
The variation of $\tau_{\rm a}$ here is applicable
only for CDA.
}
\tablenotetext{c}{
A wide range of $\beta_{\rm d}$ is investigated for CDA
whereas $\beta_{\rm d}$ is fixed at 2 for VDA.
}
\end{deluxetable}

\begin{deluxetable}{lccl}
\footnotesize  
\tablecaption{
A brief summary for the comparison between the present numerical
simulations and the latest observational results on dust and ${\rm H_2}$ properties
of galaxies. In the `consistency' column, $\bigcirc$, $\bigtriangleup$,
and $\times$ mean consistent (with observations), partly consistent (i.e.,
not all of the observations are reproduced), and inconsistent, respectively.
If no observations have been reported, then '?' is given in the consistency  column.
Observational papers (e.g.,  C12) are given in the 'Reference' column.
Clearly, a number of observations are not well reproduced by the present models,
which implies that more sophisticated numerical simulations (with better dust models)
will need to be carried out in our future studies.
\label{tbl-1}}
\tablewidth{-2pt}
\tablehead{
\colhead{  
Properties }  &
\colhead{  
Consistency  } &
\colhead{  
References } &
\colhead{  
Comments } }
\startdata
$D-A_{\rm O}$ relation
& $\bigcirc$
& G11
& Stronger SNF  and VDA are required.
\\
Large $D$ dispersion in $D-A_{\rm O}$ 
& $\times$
& G11
& New effects should be  included.
\\
$M_{\rm H_2}-M_{\rm dust}$ relation
&  $\bigtriangleup$
&  C12
& Too low $M_{\rm H_2}$ for low $M_{\rm dust}$.
\\
$f_{\rm H_2}-M_{\rm dust}$ relation
&  ?
&  $--$
& Observational test is required.
\\
$f_{\rm H_2}-M_{\ast}$ relation
& ?
& $--$
& Observational test is required.
\\
Negative radial $D$ gradient
&  $\bigcirc$
& P12
& Steeper $D$ gradients in larger $M_{\rm h}$ ?
\\
Radial $f_{\rm dust}$ gradient
&  ?
& $--$
& Different dust-to-metal ratios in ISM.
\\
Extended dusty gaseous halo
&  $\bigcirc$
&  M10
& Dust halo formation is uneversal
\\
Radial profile of dusty halo 
&  $\times$
& M10
& The simulated profiles are too steep.
\\
Halo dust in fainter galaxies
&  ?
& $--$
& Most dust can be outside optical radii.
\\
Evolution of $M_{\rm dust}$
&  $\times$
& D11
& $M_{\rm dust}$  evolution  should be  more rapid.
\\
Evolution of $M_{\rm dust}/M_{\ast}$ 
& $\times$
&  D11
& New  effects should be included.
\\
$z$-evolution of sSFR 
&  $\bigtriangleup$
&  E11
&  Lower sSFR at a given $z$ and $M_{\ast}$.
\\
\enddata
\end{deluxetable}

\clearpage
\epsscale{0.8}
\begin{figure}
\plotone{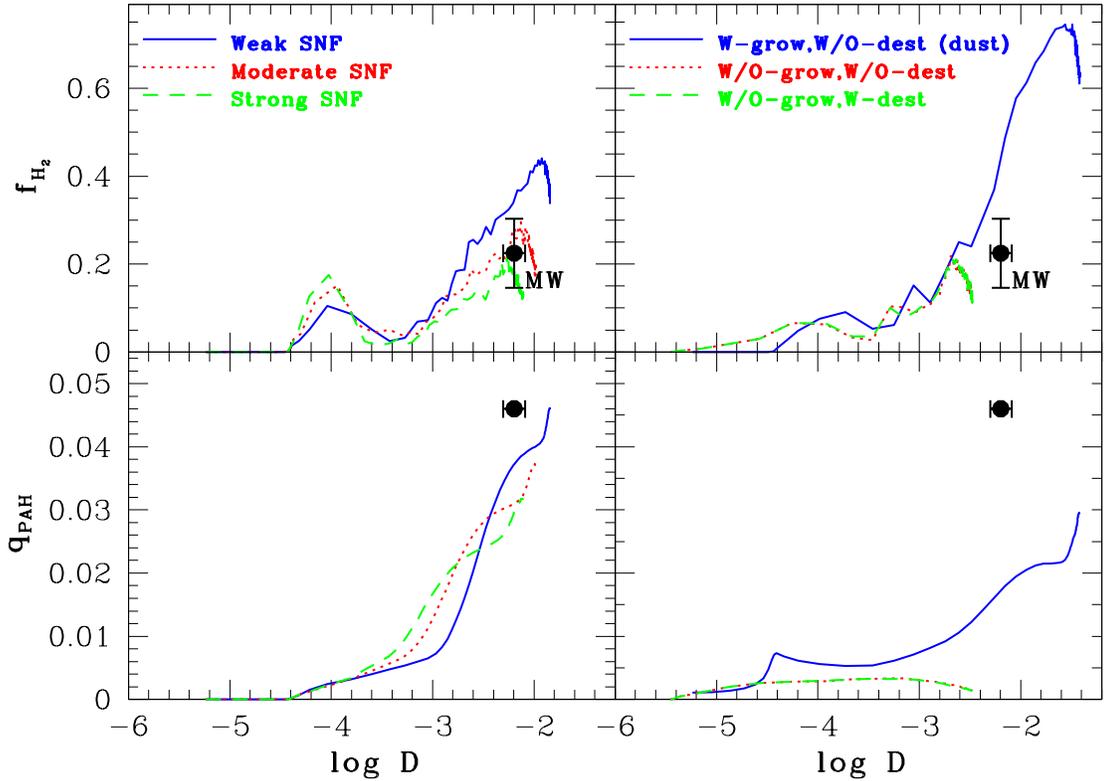}
\figcaption{
The time evolution of simulated galaxies on the $D-f_{\rm H_2}$ (upper)
and $D-q_{\rm PAH}$ planes (lower) for three different models with
different SNF (left) and those with different dust parameters (right) in
the low-resolution MW model with $M_{\rm h}=10^{12} {\it M}_{\odot}$,
$\lambda=0.06$, and CDA.
The weak, moderate, and strong SNF models are represented by
blue solid, red dotted, and green dashed lines, respectively, in the left two panels.
The models with dust growth (`W-grow') and without dust destruction ('W/O-dest'),
without dust growth ('W/O-grow') and without dust destruction,
and without dust growth and with dust destruction ('W-dest') are   represented by
blue solid, red dotted, and green dashed lines, respectively, in the right two panels.
The observational data points of the MW are shown in each panel. The observational
results from Nakanishi et al. (2003, 2006), Zubko et al. (2004)
and  Draine \& Li (2007) are adopted. The larger error bar in $f_{\rm H_2}$ is plotted
only for the purpose of showing
the uncertainly in the CO-to-${\rm H_2}$ conversion factor for the estimation
of $M_{\rm H_2}$ (thus $f_{\rm H_2}$) of the MW.
\label{fig-1}}
\end{figure}

\clearpage
\epsscale{0.8}
\begin{figure}
\plotone{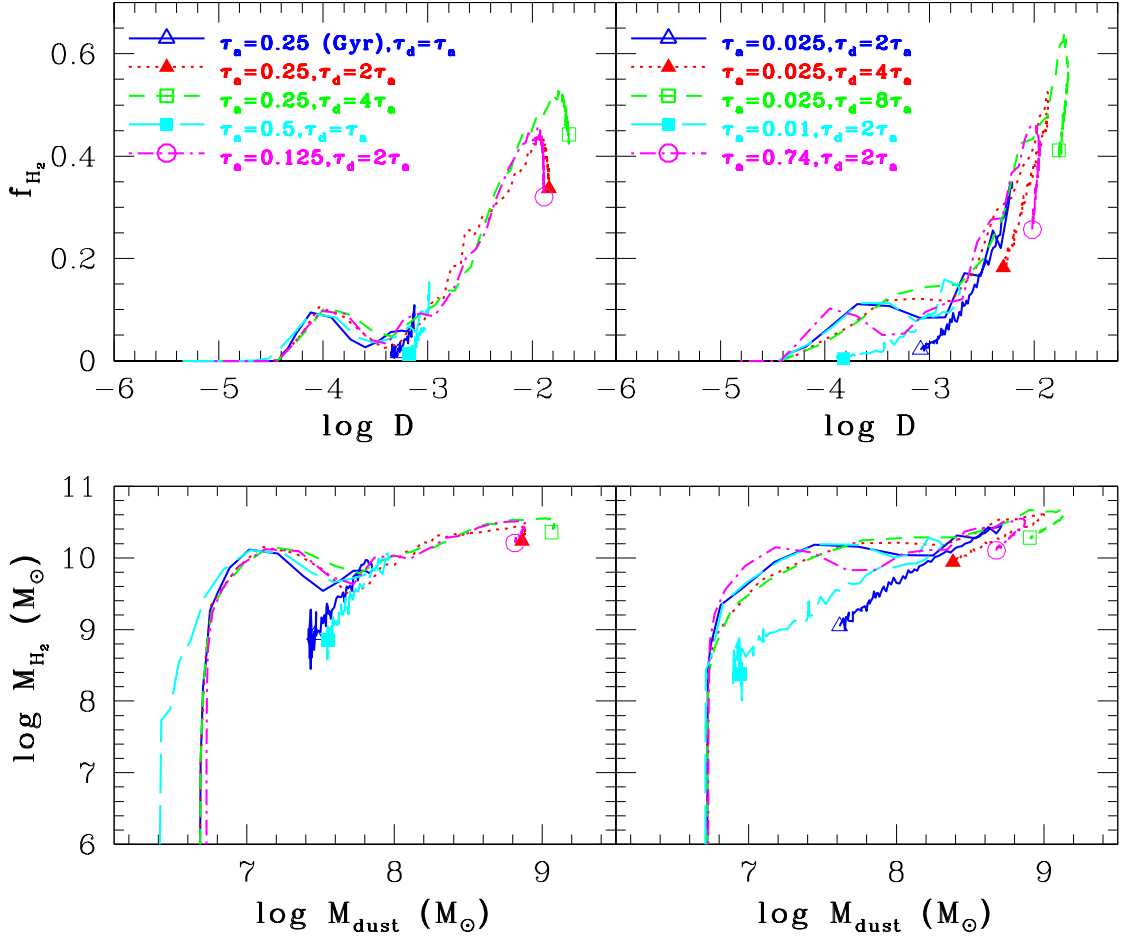}
\figcaption{
The time evolution of simulated galaxies on the $D-f_{\rm H_2}$ (upper)
and $D-M_{\rm H_2}$ planes (lower) for ten different models with
different $\tau_{\rm a}$ and $\tau_{\rm d}$
in the low-resolution MW model with $M_{\rm h}=10^{12} {\it M}_{\odot}$,$\lambda=0.06$, CDA, and weak SNF.
Different color and line-types are used for different models so that
the difference in the evolution of $f_{\rm H_2}$ and $M_{\rm dust}$ can be clearly see
n.
The final values of $f_{\rm H_2}$ and $M_{\rm H_2}$ at $z=0$
are indicated by different marks for different models.
In the left two panels, the models with $\tau_{\rm a}=\tau_{\rm d}=0.25$ Gyr
(blue solid, open triangle) andwith $\tau_{\rm a}=\tau_{\rm d}=0.5$ Gyr (cyan long-dashed, filled square)
are regarded as unrealistic owing to the later decrease of $D$ and $f_{\rm H_2}$.
In the right two panels,  the models with shorter $\tau_{\rm a}=0.025$ Gyr and 0.01 Gy
r
do not show reasonable behaviors in the evolution of $D$ and $f_{\rm H_2}$ unless
$\tau_{\rm d}$ is long enough.
These results imply that careful choice of dust growth and destruction parameters
is required for reproducing dust and ${\rm H_2}$ properties of galaxies.
\label{fig-2}}
\end{figure}

\clearpage
\epsscale{0.8}
\begin{figure}
\plotone{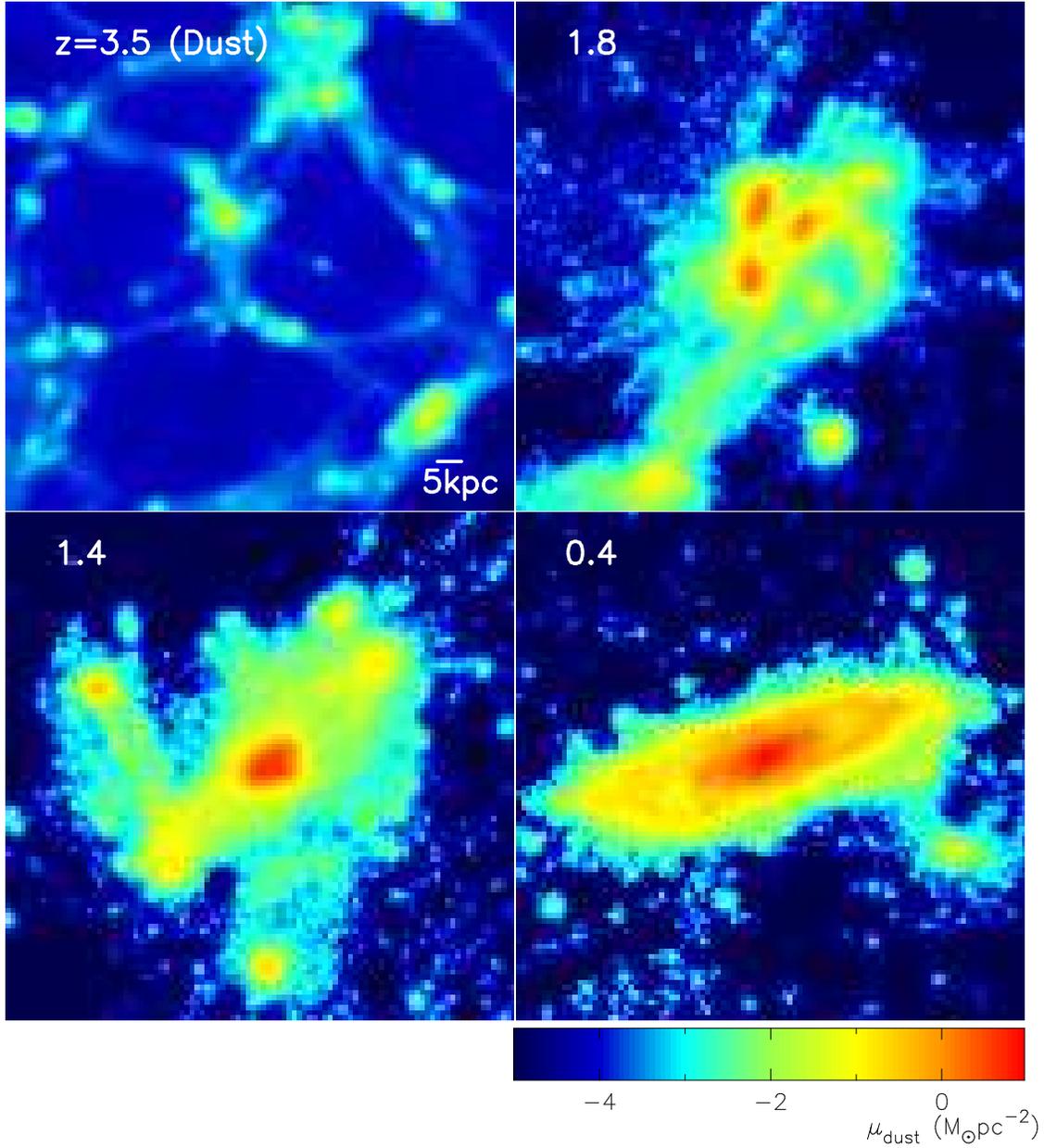}
\figcaption{
The surface mass densities of dust ($\mu_{\rm dust}$, in logarithmic scale)
projected onto the $x$-$z$ plane
at different $z$ for the fiducial (high-resolution) model
with $M_{\rm h}=10^{12} {\it M}_{\odot}$,
and $\lambda=0.06$, CDA, and the fixed Kroupa IMF.
The simulated  area is divided into $100 \times 100$ meshes so that $\mu_{\rm dust}$
can be estimated for each mesh by using the dust properties of all gas particles
in each mesh. The Gaussian smoothing with the smoothing length of 1.75 kpc is applied
for deriving a smoother  2D $\mu_{\rm dust}$  distribution for this model.
\label{fig-3}}
\end{figure}

\clearpage
\epsscale{0.8}
\begin{figure}
\plotone{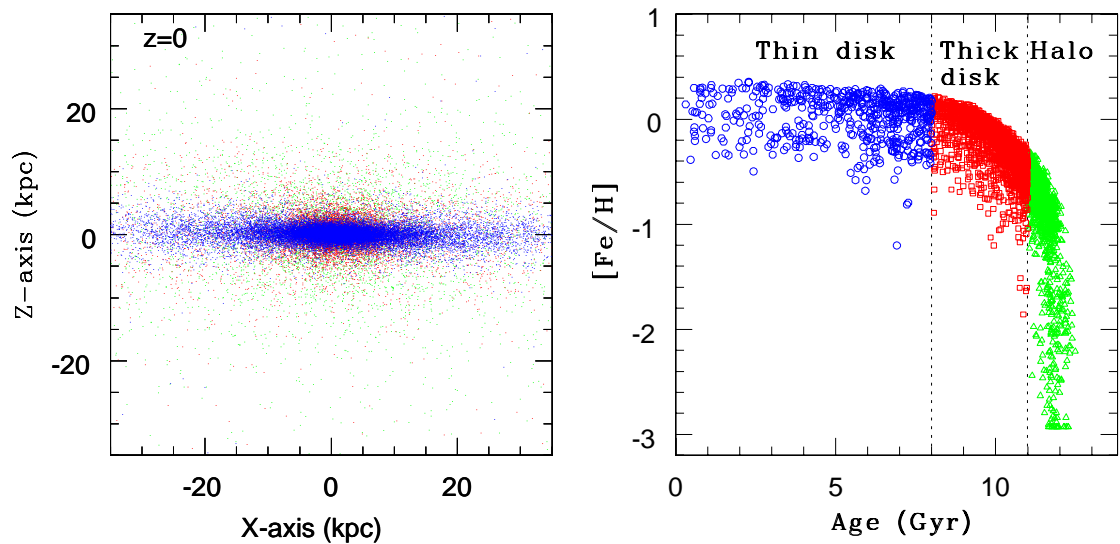}
\figcaption{
The final spatial distribution (edge-on view)
and the age-metallicity relation of new stellar particles at $z=0$
for the fiducial  model.
The new stellar particles are divided into
the thin disk (blue), thick disk (red), and stellar halo (green)
according to their ages in the low two panels. The vertical dotted lines show
the age boundary between the three components of the simulated disk galaxy.
Only one every 20 particles for the left and one every 100 particles
for the right are  shown so that the file size of this figure can be
kept small enough (less than Mega byte).
\label{fig-4}}
\end{figure}

\clearpage
\epsscale{0.8}
\begin{figure}
\plotone{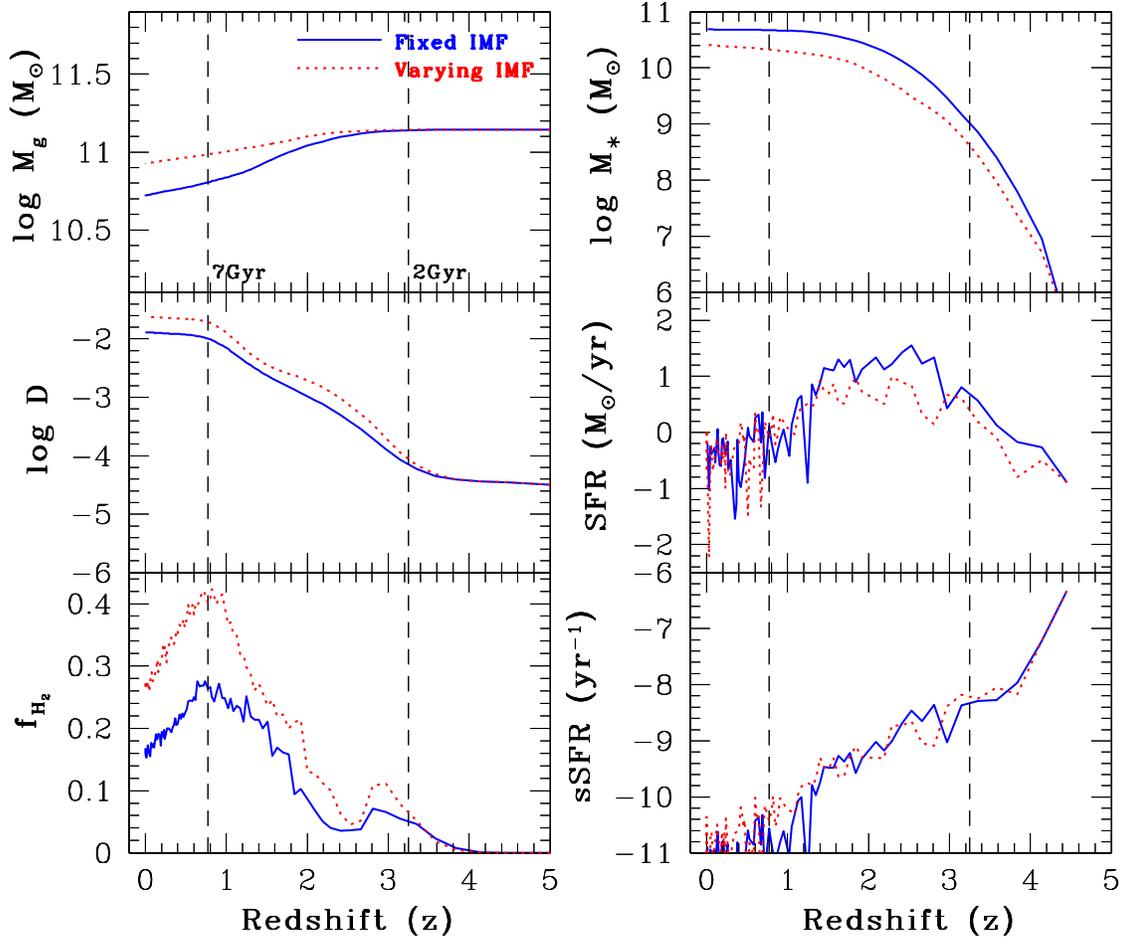}
\figcaption{
The left three panels show
the $z$-evolution of $M_{\rm g}$ (top), $\log D$ (middle),
and $f_{\rm H_2}$ (bottom)
in the fiducial
model for the fixed Kroupa IMF (blue solid) and the time-varying IMF (red dotted).
The  right three show
the $z$-evolution of $M_{\ast}$ (top), SFR (middle),
and sSFR (specific SFR, SFR$/M_{\ast}$, bottom) for the model.
The two vertical dashed lines indicate the ages of the universe at the two $z$.
The slow steady increase of $M_{\rm dust}$ and $D$ at $z>3.5$ is due to the  growth of
pre-galactic dust whereas the rapid increase at $z \le 3.5$
 is caused by  chemical enrichment
due to very active star formation during early disk formation via dissipative merging
of subgalactic clumps.
\label{fig-5}}
\end{figure}

\clearpage
\epsscale{1.0}
\begin{figure}
\plotone{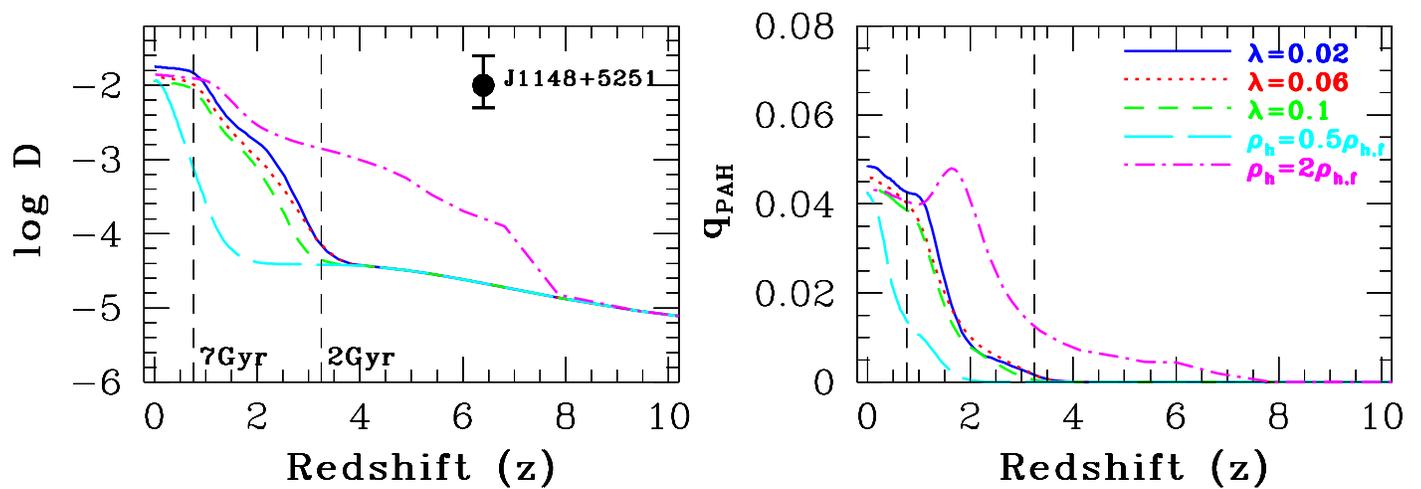}
\figcaption{
The $z$-evolution of $\log D$ (left),
and $q_{\rm PAH}$ (right) in the models
with $\lambda=0.02$ (blue solid),
$\lambda=0.06$ (red dotted),
$\lambda=0.1$ (green short-dashed),
$\rho_{\rm h}=0.5\rho_{\rm h, f}$ (cyan long-dashed),
and $\rho_{\rm h}=2\rho_{\rm h, f}$ (magenta dot-dashed).
The two vertical dashed lines indicate the ages of the universe at the two $z$.
For comparison, the observational result of high-$z$ QSO SDSS J1148+5251 is shown
by a filled black circle with observationally and theoretically inferred error bars.
\label{fig-6}}
\end{figure}

\clearpage
\epsscale{0.8}
\begin{figure}
\plotone{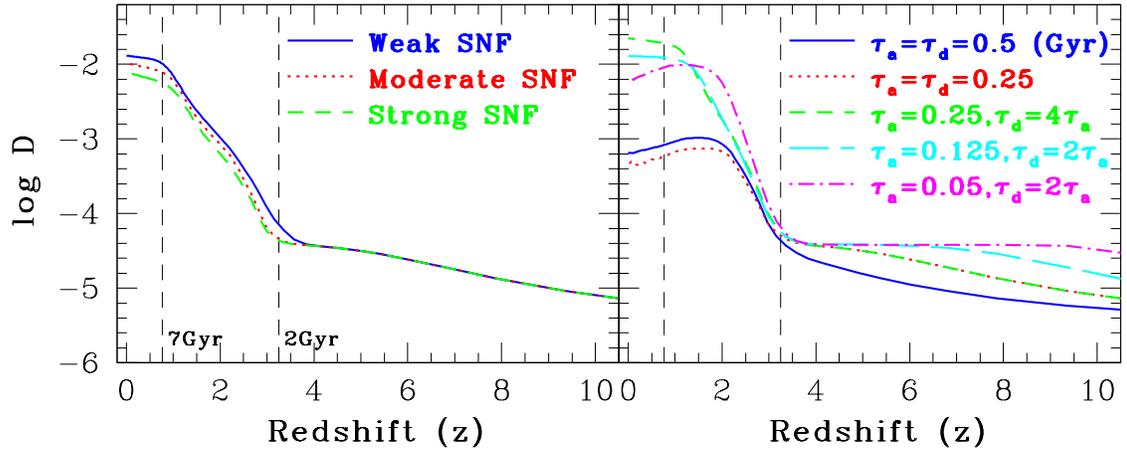}
\figcaption{
The $z$-evolution of $\log D$ in the simulated galaxies
with weak (blue solid),
moderate (red dotted),
and strong SNF (green short-dashed) in the left panel
and with $\tau_{\rm a}=\tau_{\rm d}=0.5$ Gyr (blue solid),
$\tau_{\rm a}=\tau_{\rm d}=0.25$   (red dotted),
$\tau_{\rm a}=0.25$ Gyr $\tau_{\rm d}=4\tau_{\rm d}$   (green short-dashed),
$\tau_{\rm a}=0.125$ Gyr and $\tau_{\rm d}=2\tau_{\rm a}$  (cyan long-dashed),
and $\tau_{\rm a}=0.05$ Gyr and $ \tau_{\rm d}=2\tau_{\rm a}$  (magenta dot-dashed).
\label{fig-7}}
\end{figure}

\clearpage
\epsscale{0.8}
\begin{figure}
\plotone{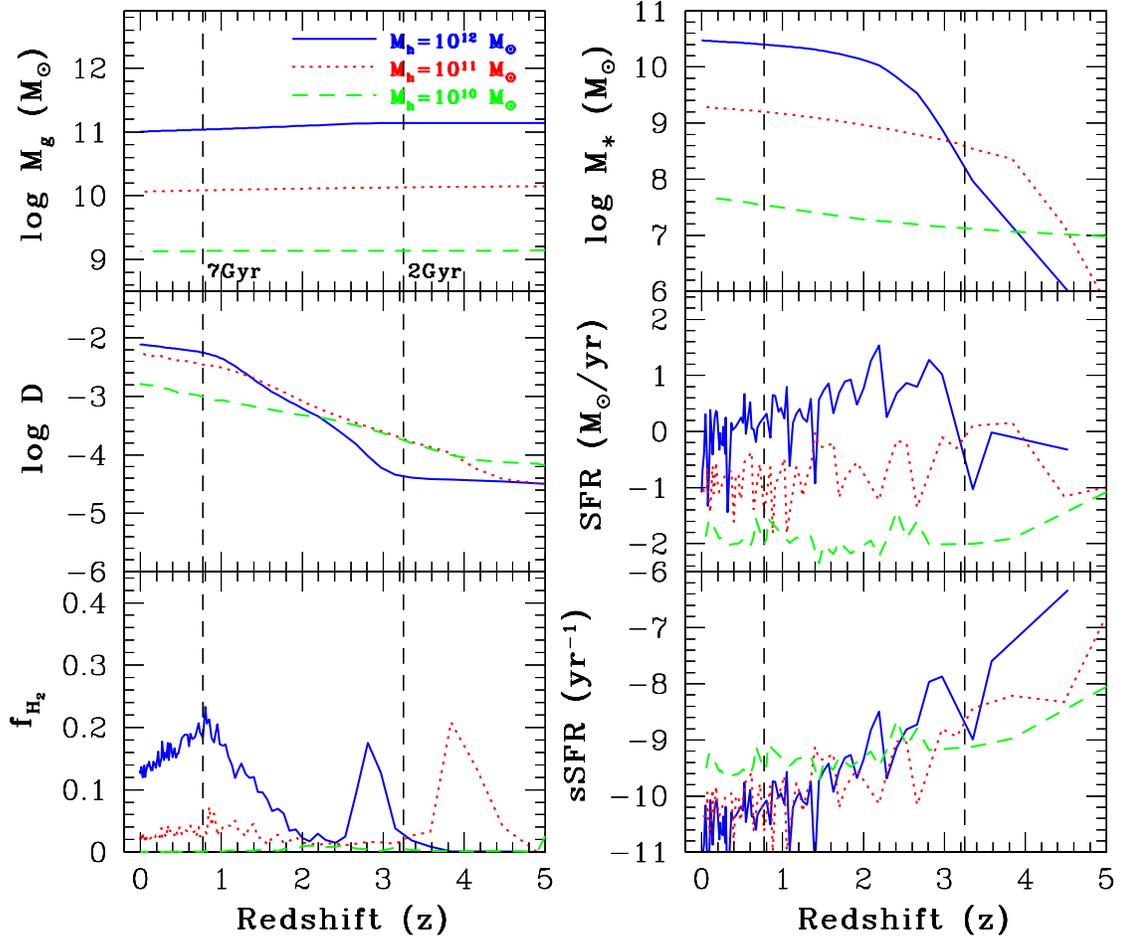}
\figcaption{
The same as Figure 5 but for the models with CDA, strong
SNF,  and different $M_{\rm h}$:
$M_{\rm h}=10^{12} {\it M}_{\odot}$ (blue solid),
$M_{\rm h}=10^{11} {\it M}_{\odot}$ (red dotted),
and $M_{\rm h}=10^{10} {\it M}_{\odot}$ (green short-dashed).
The spin parameter is fixed at $\lambda=0.06$ for the three models.
\label{fig-8}}
\end{figure}

\clearpage
\epsscale{0.8}
\begin{figure}
\plotone{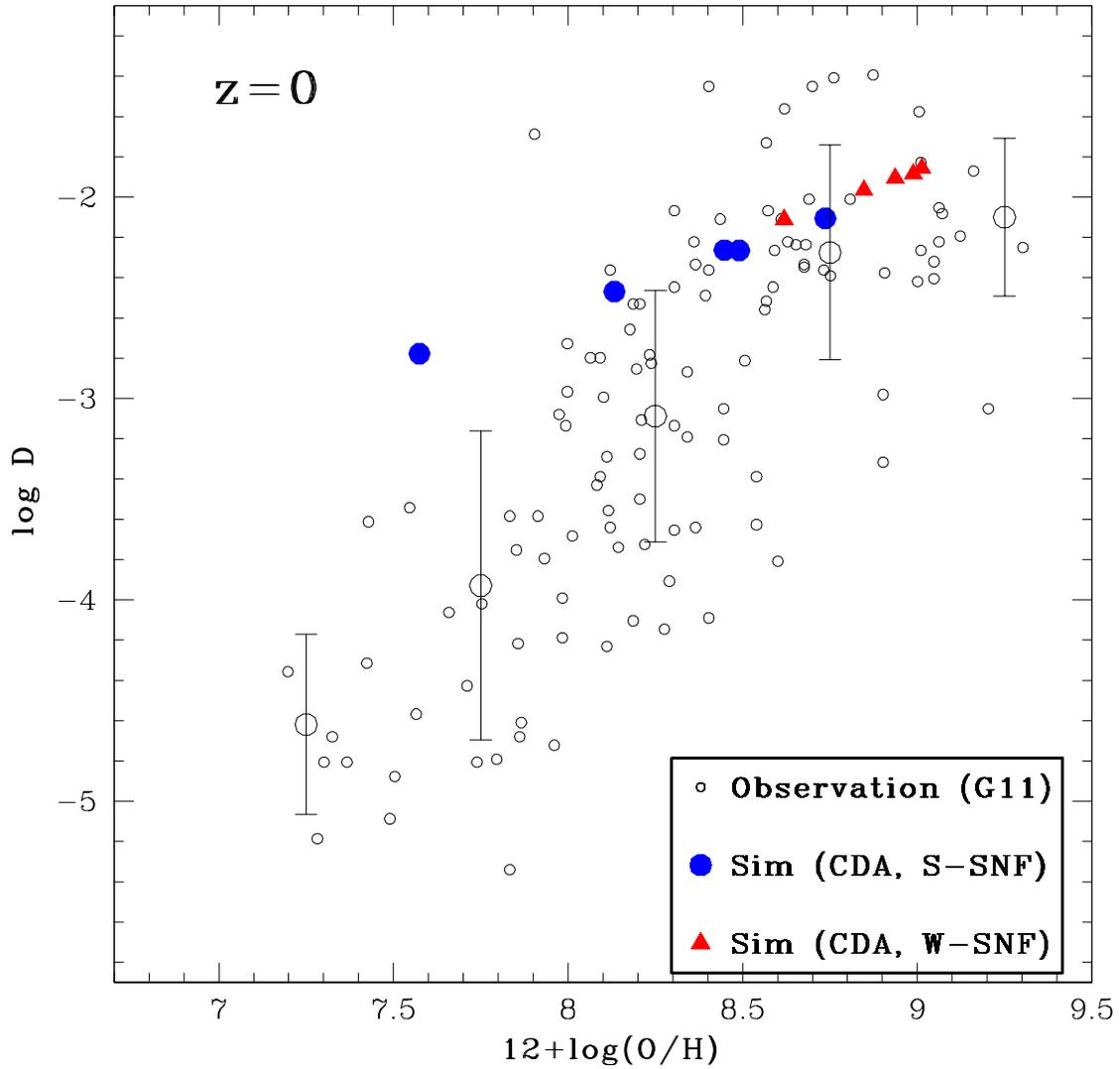}
\figcaption{
The metallicity-dust relation for the models with CDA at $z=0$.
The locations of five simulated galaxies with initially different $M_{\rm h}$
on the $A_{\rm O}-D$ plane is shown for weak SNF ('W-SNF', filled red triangles)
and for strong SNF ('S-SNF', filled blue circles).
The observational data from Galametz et al. (2011; G11) are shown by open black circles.
The mean values and dispersions of the observational data at five bins
are given by  big open black circles and error-bars.
\label{fig-9}}
\end{figure}

\clearpage
\epsscale{0.8}
\begin{figure}
\plotone{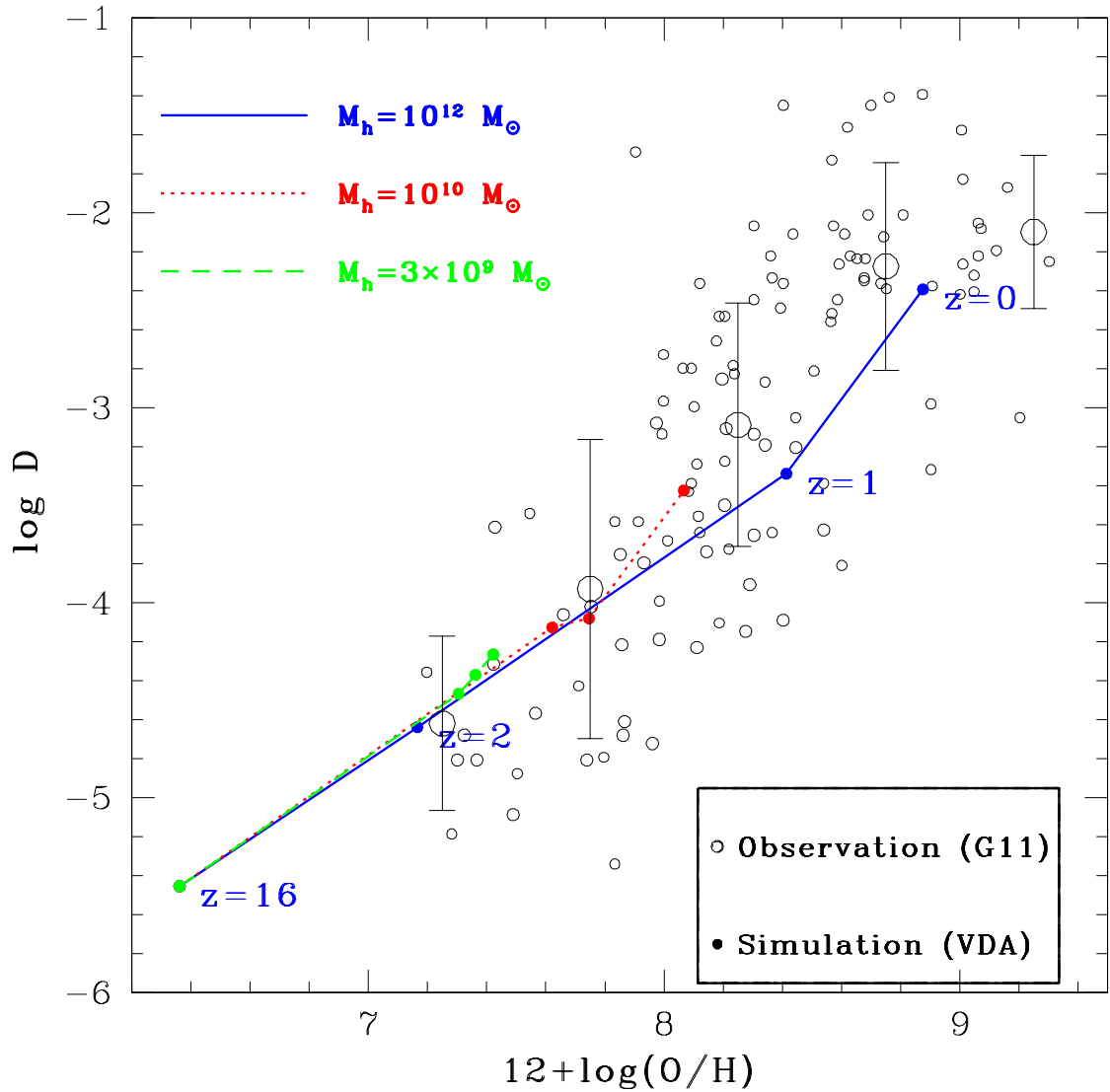}
\figcaption{
The same as Figure 8 but for the three different $M_{\rm h}$
models with strong SNF and VDA at  $z=0$, 1, 2, and 16
(i.e., starting redshift, $z_{\rm i}$).
The models with
${\it M}_{\rm h}=10^{12} {\it M}_{\odot}$,
${\it M}_{\rm h}=10^{11} {\it M}_{\odot}$,
and ${\it M}_{\rm h}=3\times 10^{9} {\it M}_{\odot}$
are shown by blue solid, red dotted, and green short-dashed lines, respectively.
\label{fig-10}}
\end{figure}

\clearpage
\epsscale{1.0}
\begin{figure}
\plotone{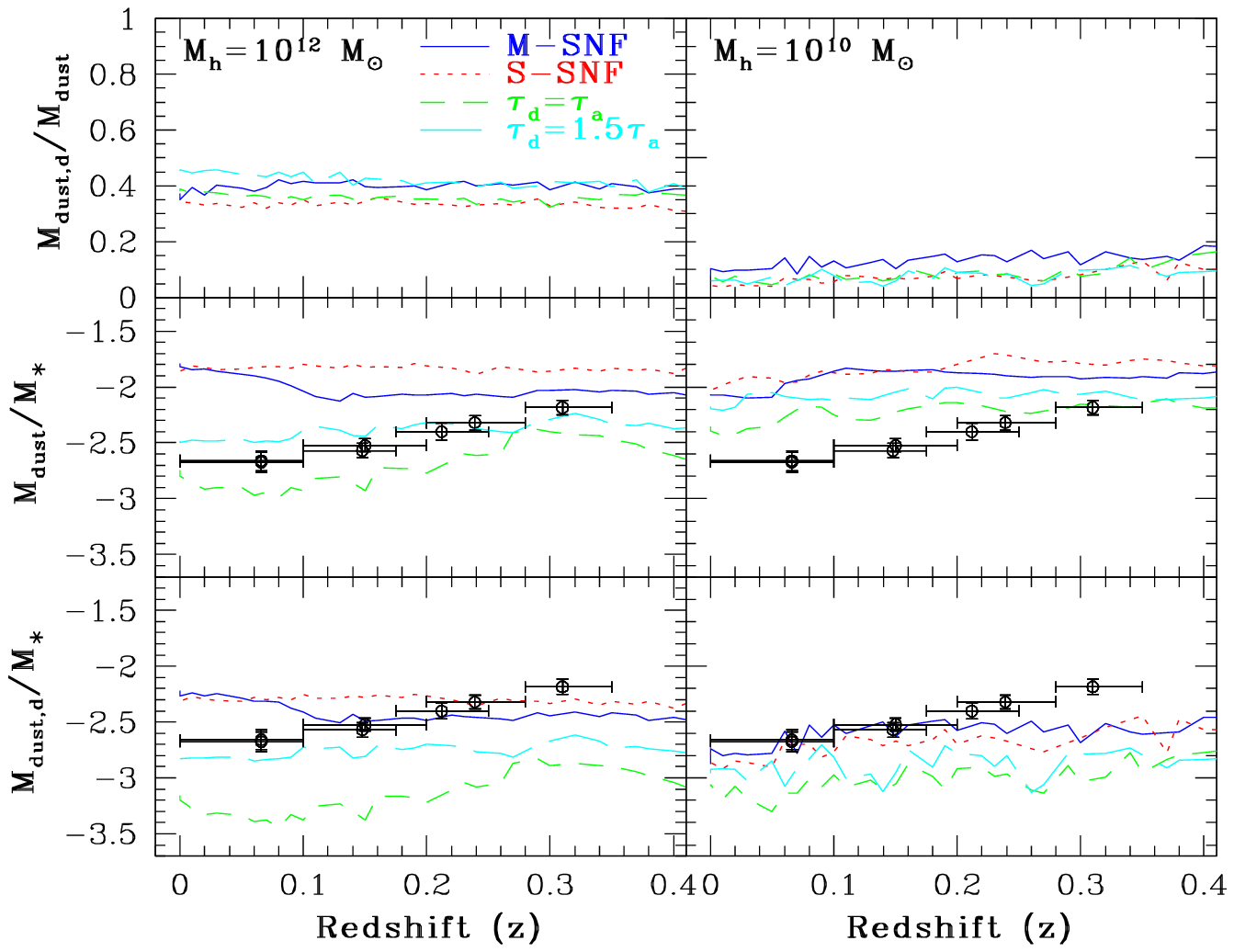}
\figcaption{
The $z$-evolution of
$M_{\rm dust, d}/M_{\rm dust}$ (top),
$M_{\rm dust}/M_{\ast}$ (middle),
and  $M_{\rm dust, d}/M_{\ast}$ (bottom)
for the two representative sets of
models with VDA (different $\beta_{\rm d}$) and $\lambda=0.06$.
The left and right panels are  for $M_{\rm h}=10^{12} {\it M}_{\odot}$
$M_{\rm h}=10^{12} {\it M}_{\odot}$ and
and $M_{\rm h}=10^{10} {\it M}_{\odot}$, respectively.
strong SNF, and standard VDA with $\tau_{\rm d}=2\tau_{\rm a}$ (blue solid),
Different colors (and line-types) shows the models with different SNF and $\beta_{\rm d}$:
strong SNF and $\tau_{\rm d}=2\tau_{\rm a}$ (blue solid),
moderate SNF and $\tau_{\rm d}=2\tau_{\rm a}$ (red dotted),
strong SNF and $\tau_{\rm d}=\tau_{\rm a}$ (green short-dashed)
and strong SNF and $\tau_{\rm d}=1.5\tau_{\rm a}$ (cyan long-dashed).
The observational results from D11 are shown for comparison.
\label{fig-11}}
\end{figure}

\clearpage
\epsscale{0.8}
\begin{figure}
\plotone{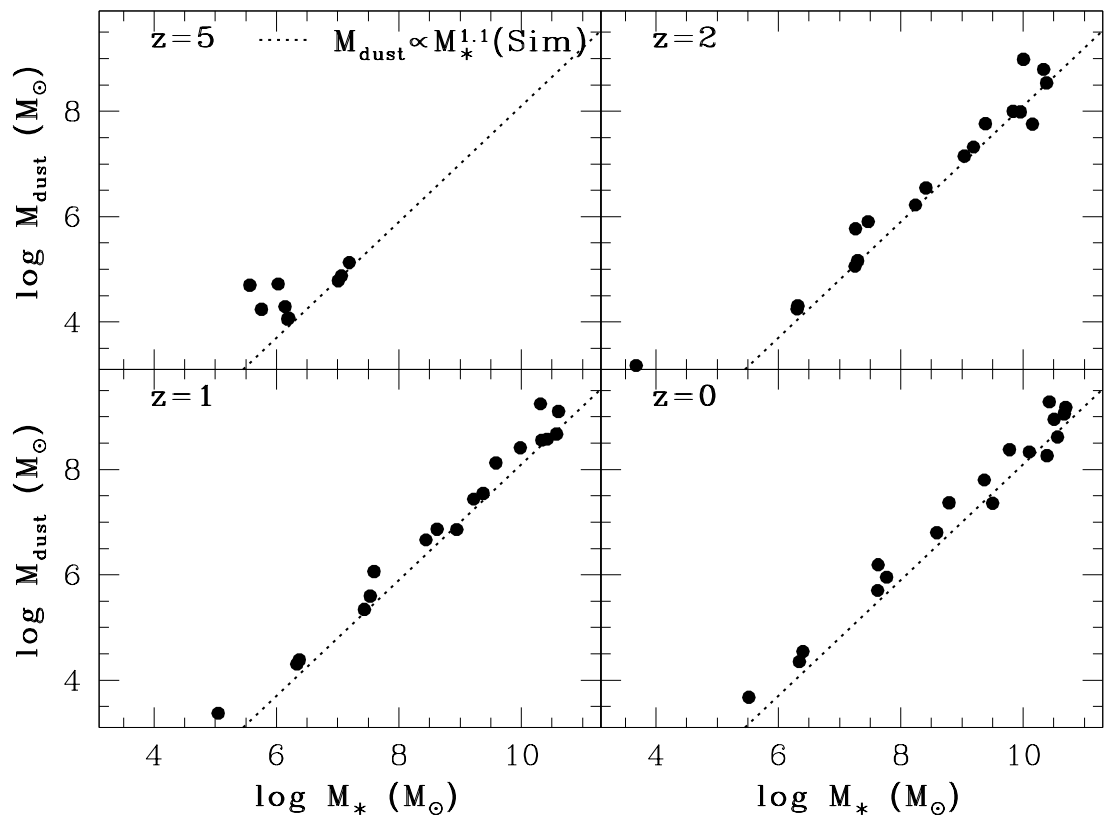}
\figcaption{
The locations of the eighteen simulated galaxies in the models with VDA
on the $M_{\ast}-M_{\rm dust}$ plane
at $z=5$ (upper left), 2 (upper right), 1 (lower left), and 0 (lower right).
A scaling relation
of $M_{\rm dust} \propto M_{\ast}^{1.1}$ that appears to fit with the simulated data
at $z=0$
is shown by a dotted line in each panel just for comparison.
The simulated galaxies with different $M_{\rm h}$, $\lambda$, and $\rho_{\rm h}$
are shown here.
\label{fig-12}}
\end{figure}

\clearpage
\epsscale{0.8}
\begin{figure}
\plotone{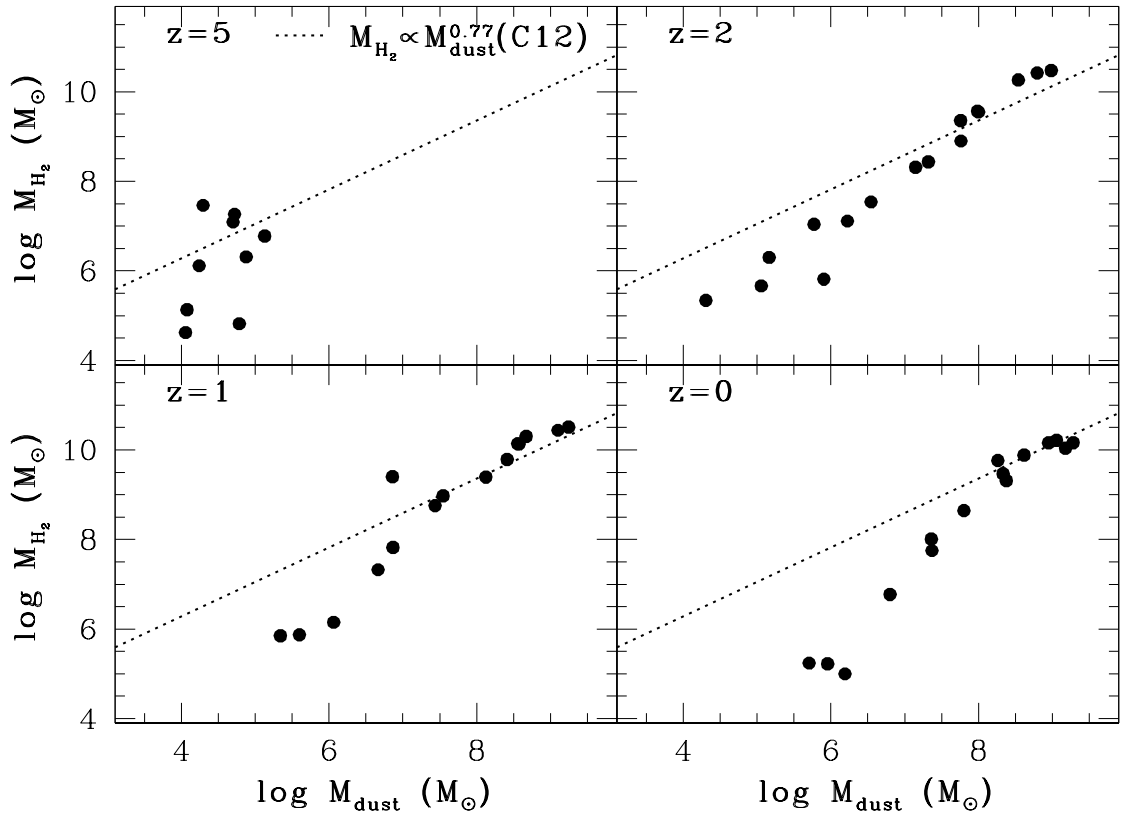}
\figcaption{
The same as Figure 12 but for
the $M_{\rm dust }-M_{\rm H_2}$ relation.
The observed dust scaling relation
of $M_{\rm dust} \propto M_{\ast}^{0.77}$ derived by C12
is shown by a dotted line in each panel just for comparison.
\label{fig-13}}
\end{figure}

\clearpage
\epsscale{0.8}
\begin{figure}
\plotone{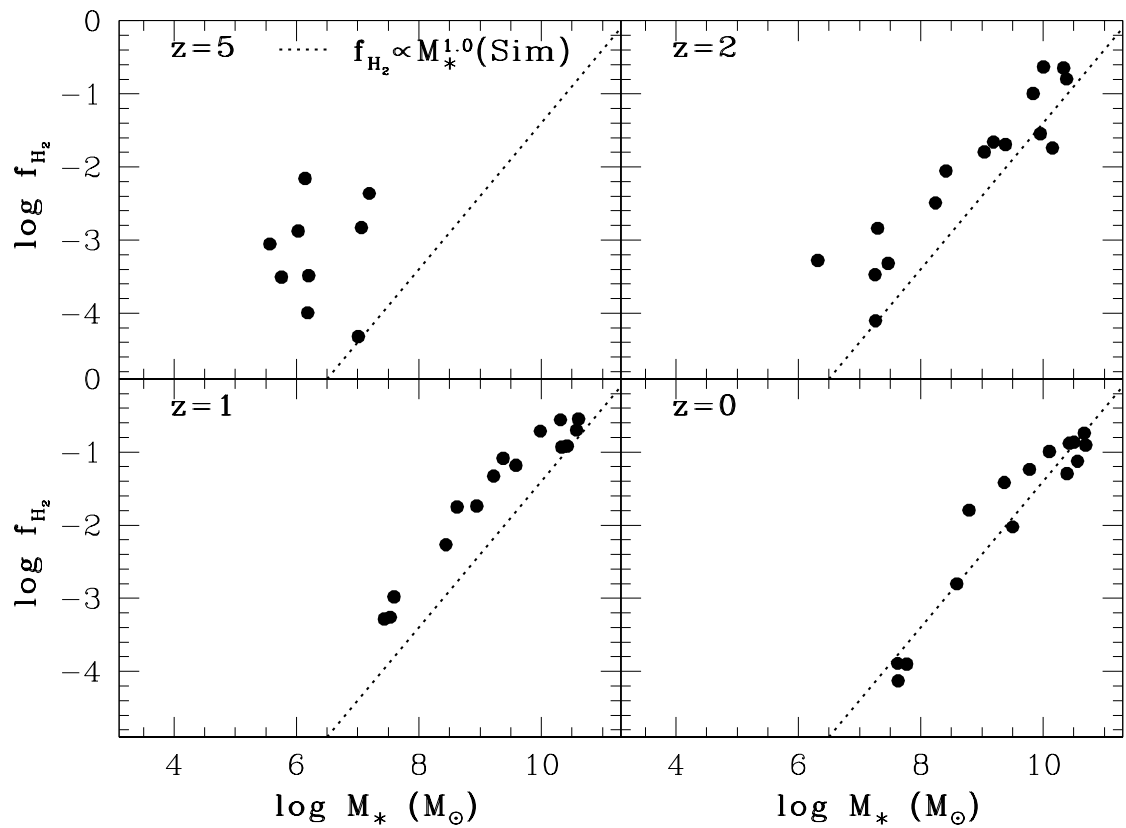}
\figcaption{
The same as Figure 12 but for
the $M_{\ast}-f_{\rm H_2}$ relation.
A scaling relation
of $f_{\rm H_2} \propto M_{\ast}^{1.0}$ that appears to fit with the simulated data
points at $z=0$
is shown by a dotted line in each panel just for comparison.
\label{fig-14}}
\end{figure}

\clearpage
\epsscale{0.8}
\begin{figure}
\plotone{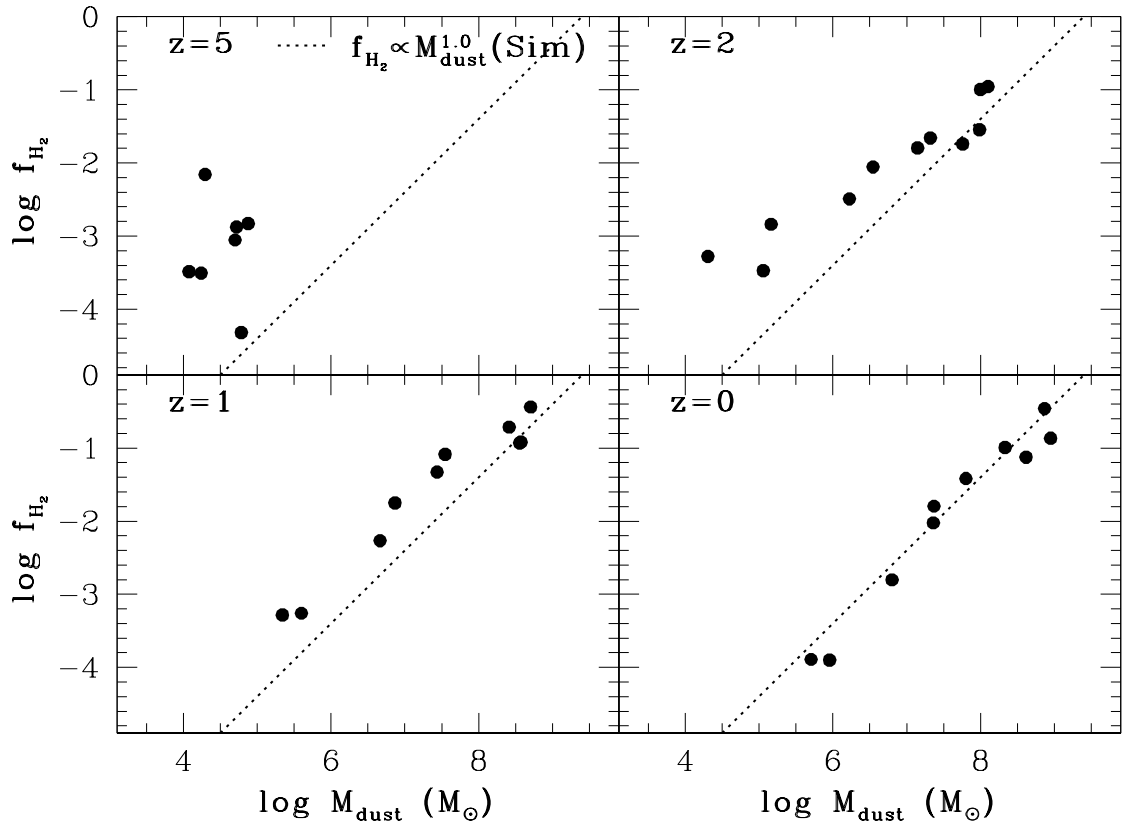}
\figcaption{
The same as Figure 12 but for
the $M_{\rm dust}-f_{\rm H_2}$ relation.
A scaling relation
of $f_{\rm H_2} \propto M_{\rm dust}^{1.0}$
is shown by a dotted line in each panel just for comparison.
\label{fig-15}}
\end{figure}

\clearpage
\epsscale{0.8}
\begin{figure}
\plotone{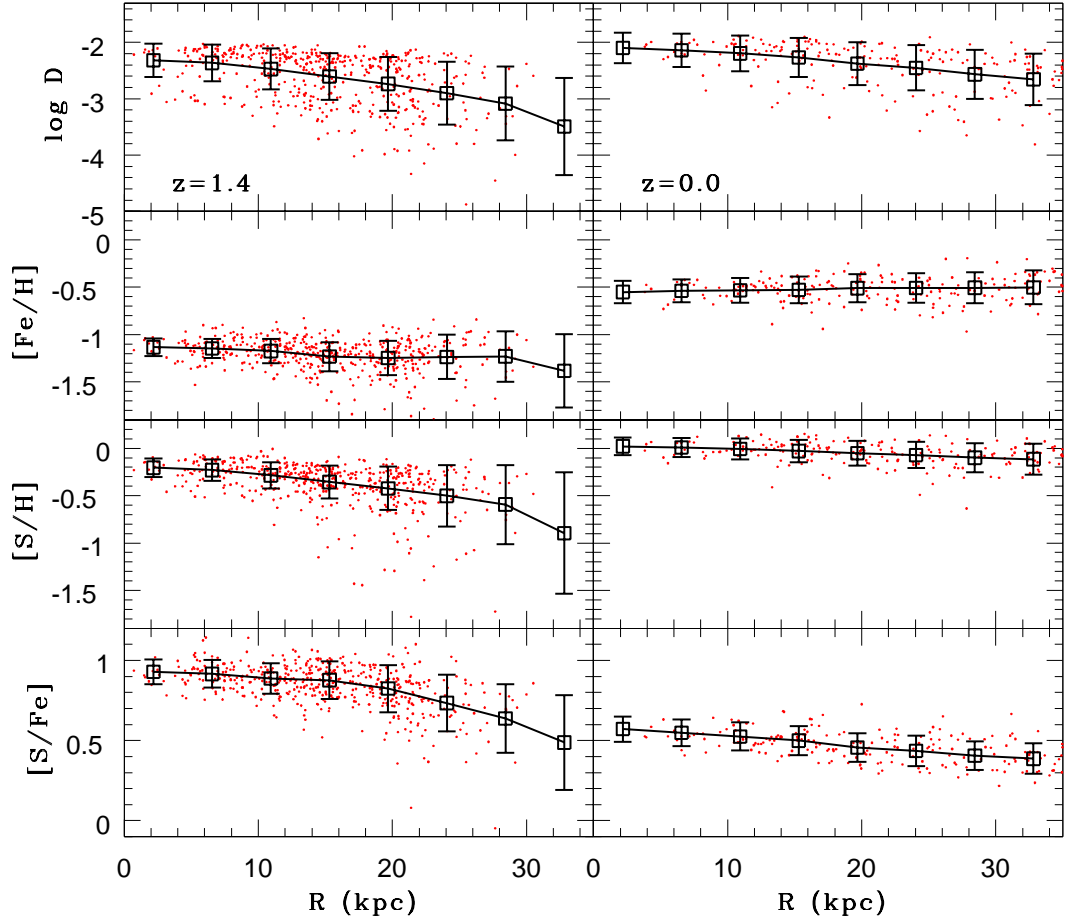}
\figcaption{
The radial gradients of $D$ (top),  [Fe/H] (second from the top), [S/H] (second from
the bottom), and [S/Fe] (bottom) of the gas disk at $z=1.4$ (left) and 0 (right)
for the model in which (i) initial conditions are exactly the same as those in the fiducial
MW model and (ii)  VDA instead of CDA is adopted.
The black solid line connects mean values at different radial ($R$) bins shown
by open squares and the error bars indicate the dispersions of these four physical
quantifies.  The small red dots represent $D$  and gas-phase
chemical abundances for selected
gas particles (corresponding to individual local gaseous regions).
The dust depletion level can be measured from [S/Fe] in the present study where only
S is assumed to be unable to be locked up into dust grains.
Accordingly, higher [S/Fe] means a higher level of dust depletion in this figure.
\label{fig-16}}
\end{figure}

\clearpage
\epsscale{0.8}
\begin{figure}
\plotone{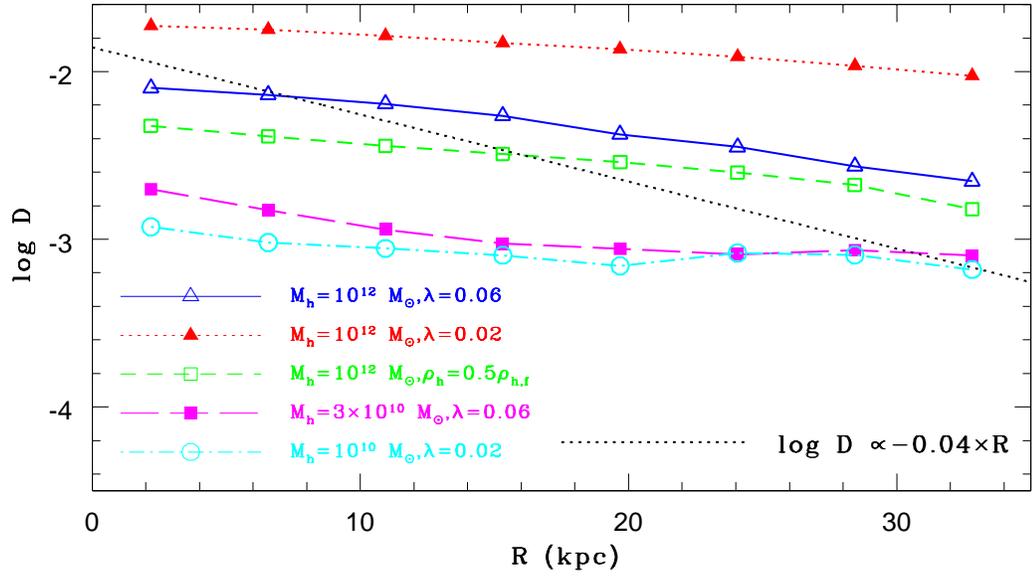}
\figcaption{
The radial gradients of $D$ (upper) and [S/Fe] (lower) at $z=0$ for five
different models
with $M_{\rm h}=10^{12} {\it M}_{\odot}$ and $\lambda=0.06$ (blue solid),
$M_{\rm h}=10^{12} {\it M}_{\odot}$ and $\lambda=0.02$ (red dotted),
$M_{\rm h}=10^{12} {\it M}_{\odot}$ and $\rho_{\rm h}=0.5 \rho_{\rm h,f}$
(green short-dashed),
$M_{\rm h}=3 \times 10^{10} {\it M}_{\odot}$ and
$\lambda=0.06$ (cyan long-dashed),
and $M_{\rm h}=10^{10} {\it M}_{\odot}$ and $\lambda=0.02$ (magenta dot-dashed).
In these models, VDA is adopted.
For comparison, the radial gradient of $\log D \propto -0.04R$ is shown by a
black dotted line for comparison.
\label{fig-17}}
\end{figure}

\clearpage
\epsscale{0.8}
\begin{figure}
\plotone{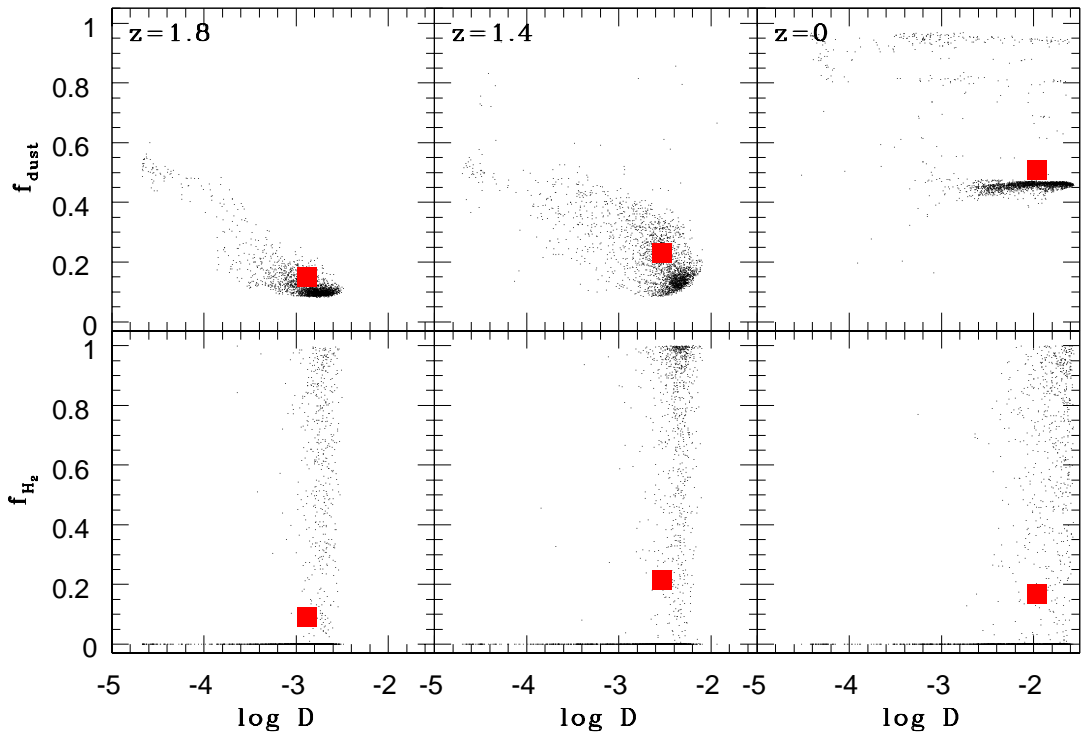}
\figcaption{
The locations of the selected gas particles of the simulated galaxy in the fiducial model
on the $D-f_{\rm dust}$ (upper) and $D-f_{\rm H_2}$
(lower) planes at $z=1.8$ (let), 1.4 (middle),
and 0 (right).  Each small black dot
represents each gas particle in the three panels, and the large red square represents
the mean $f_{\rm dust}$ and $f_{\rm H_2}$ in each panel.
The file size of this figure can be very large ($\sim 10$ MB), if all gas particles
are plotted. Therefore, one every 50 particles is plotted
so that the file size can be much smaller.
Nevertheless the simulated $D-f_{\rm dust}$ and $D-f_{\rm H_2}$ can be
clearly seen.
The dust-to-metal ratios ($f_{\rm dust}$)
are quite different between different gas particles
and such differences can be more clearly seen in higher $z$.
The mean $f_{\rm dust}$ is larger at higher in this model,
and there is a clear concentration of gas particles around $\log D \sim -2$ and
$f_{\rm dust} \sim 0.4$ at $z=0$.
\label{fig-18}}
\end{figure}

\clearpage
\epsscale{0.8}
\begin{figure}
\plotone{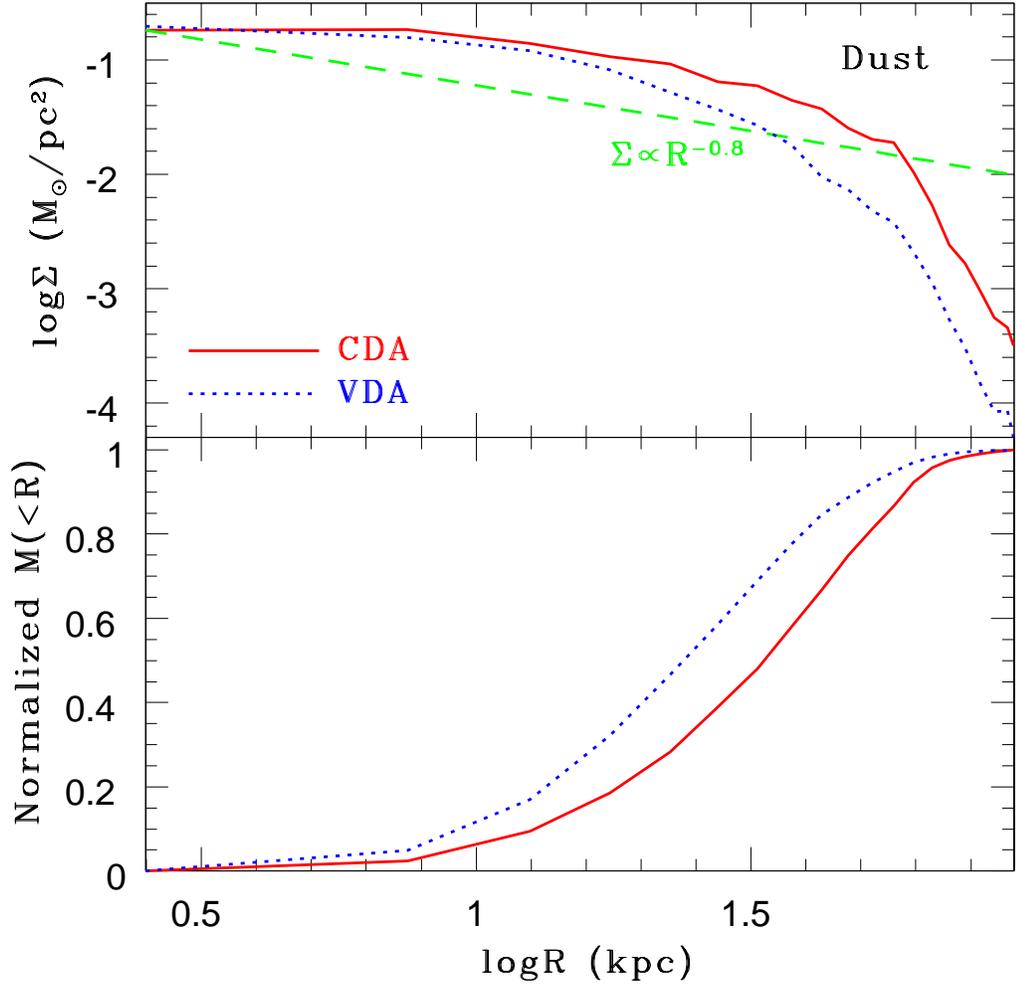}
\figcaption{
The projected mass profile ($\Sigma$, upper) and (normalized) cumulative
mass distribution ($M(<R)$, lower) for dust
in the MW-like models with $M_{\rm h}=10^{12} {\it M}_{\odot}$
and  $\lambda=0.06$
for CDA (red solid) and  VDA (blue dotted).
For comparison, the observed slope ($\Sigma \propto R^{-0.8}$) from
M\'enard et al. (2010, M10) is shown by a green dashed line.
Clearly, the simulated profile of the extended dusty halo is much steeper
than the observed one.
Also a large fraction of dust can be in the outer halo for these model
($M(R<30$kpc)/$M(R<100$kpc)$ > 0.5$).
\label{fig-19}}
\end{figure}

\clearpage
\epsscale{1.0}
\begin{figure}
\plotone{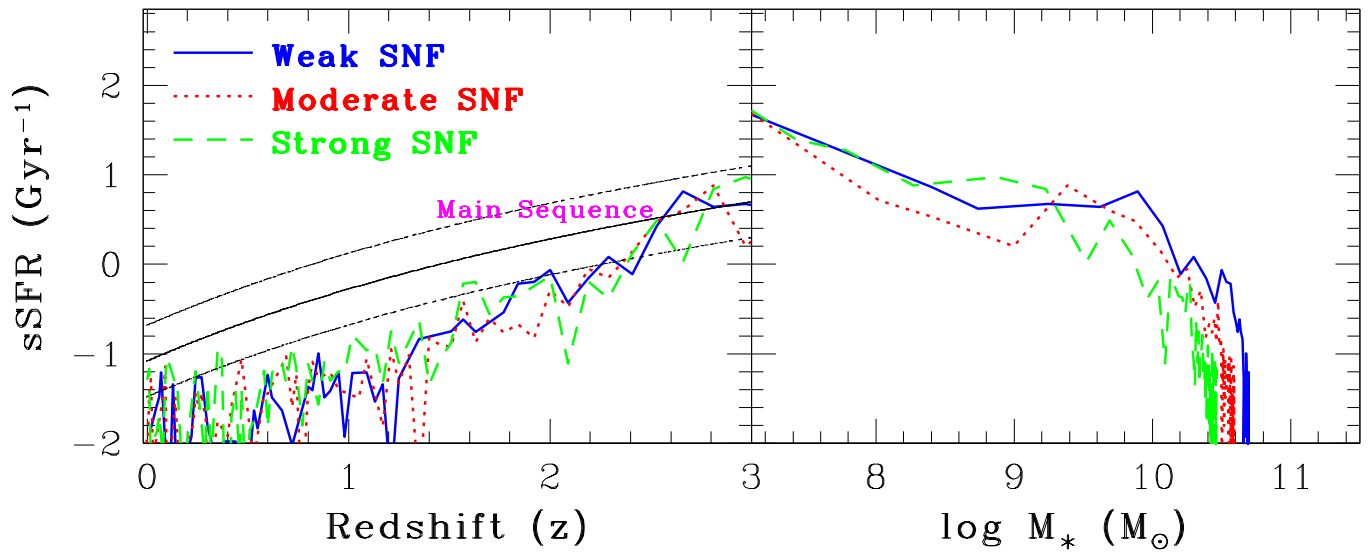}
\figcaption{
The evolution of specific star formation rate (sSFR=SFR$/M_{\ast}$) as
a function of $z$ (left) and $M_{\ast}$ (right)
for the MW-like models with $M_{\rm h}=10^{12} {\it M}_{\odot}$, $\lambda=0.06$,
and VDA.
The blue solid, red dotted, and green short-dashed lines represent the models
with weak,  moderate, and strong SNF, respectively.
 The observed relation
for the main sequence galaxies (Elbaz et al. 2011; E11) is shown  by a solid line
and the observed range of sSFR at a given $z$
in the relation is indicated by two dotted lines.
\label{fig-20}}
\end{figure}

\end{document}